%% file: main_tr.tex
\documentclass{article}
\usepackage[preprint,nonatbib]{neurips_2024}

\usepackage[T1]{fontenc}
\usepackage[utf8]{inputenc}
\usepackage{lmodern}  % NeurIPS 2024 no longer auto-loads Times; swap to mathptmx for a Times look
\usepackage{microtype}
\usepackage{indentfirst}        % also indent the first paragraph after each heading
\makeatletter
\@ifundefined{@noticestring}{}{\renewcommand{\@noticestring}{Preprint.}}  % arXiv: drop "Under review."
\makeatother

\usepackage{amsmath}
\usepackage{amssymb}
\usepackage{amsthm}
\numberwithin{equation}{section}  % per-section equation numbering (e.g. 2.1, 2.3)

\usepackage{booktabs}
\usepackage{array}
\usepackage{multirow}
\usepackage{graphicx}
\graphicspath{{figures/}}
\usepackage{caption}
\usepackage{xcolor}
\newcommand{\fitwidth}[1]{\resizebox{\ifdim\width>\linewidth \linewidth\else\width\fi}{!}{#1}}

\usepackage[section]{placeins}

\usepackage{enumitem}
\usepackage[natbibapa]{apacite}
\definecolor{linkblue}{RGB}{0,68,170}
\usepackage[colorlinks=true,citecolor=linkblue,urlcolor=linkblue,linkcolor=linkblue,
            breaklinks=true]{hyperref}
\providecommand{\doi}[1]{\href{https://doi.org/#1}{#1}}

\usepackage{algorithm}
\usepackage{algpseudocode}
\algrenewcommand\algorithmicrequire{\textbf{Input:}}
\algrenewcommand\algorithmicensure{\textbf{Output:}}
\floatname{algorithm}{Procedure}

\theoremstyle{definition}
\newtheorem{definition}{Definition}

\newcommand{\bten}{b_{10}}
\newcommand{\bzo}{b_{01}}
\newcommand{\armP}{\textsf{P}}
\newcommand{\armB}{\textsf{B}}
\newcommand{\armTC}{\textsf{TC}}
\newcommand{\armCF}{\textsf{CF-SF}}
\newcommand{\armGR}{\textsf{GR-SHAPE}}
\newcommand{\armBI}{\textsf{BI}}
\newcommand{\Cone}{C1$'$}
\newcommand{\Ctwo}{C2$'$}
\newcommand{\Cthree}{C3$'$}
\newcommand{\Ipub}{\ensuremath{\mathcal{I}_{\mathrm{base}}}}
\newcommand{\Ihid}{\ensuremath{\mathcal{I}_{\mathrm{plus}}}}

\title{\textbf{Falsification, Not Exposure: An Internally Preregistered Placebo-Controlled Decomposition of Self-Repair Feedback in Frozen Small Code Models}}

\author{Mehmet \.{I}\c{s}can\thanks{Corresponding author. PythaLab, Y{\i}ld{\i}z Technical
University, Istanbul, Turkey. E-mail: \href{mailto:miscan@yildiz.edu.tr}{\texttt{miscan@yildiz.edu.tr}}.}\\
\small PythaLab, Y{\i}ld{\i}z Technical University, Istanbul, Turkey}
\date{}

\begin{document}
\maketitle

\begin{abstract}
\noindent
In deployment settings where retraining is not feasible, small \emph{frozen code models} are commonly asked to repair a failed program 
after being shown their own failing output. This practice is often treated as an additional \emph{retry mechanism}.
From a Popperian perspective, however, it raises a deeper epistemic and measurement problem. 
A generated program can be viewed as a tentative \emph{conjecture}, while a 
violation observed during test execution can be viewed as a test-oracle-relative 
and executable \emph{counterexample}. The scientific value of feedback should therefore 
be attributed not to re-exposure to failing code itself, but to whether the candidate 
conjecture is opened to external, executable, and auditable criticism. Prior work has often bundled failing code, 
execution evidence, and retry instruction into a single \emph{feedback packet}, without decomposing packet components 
against a \emph{blind resampling} baseline under the same \emph{output-generation budget} and against content-free 
placebo controls. As the third stage of a falsification-centered measurement program, this study presents a 
placebo-controlled \emph{evaluation instrument} that compares bare failing code, executed facts, code-plus-facts, 
neither, and template- and bullet-count-matched placebo conditions on the same task--model units. The contribution 
is not a new \emph{repair algorithm}, but a reflexive measurement methodology combining \emph{packet decomposition}, \emph{placebo mirroring}, 
matched-budget comparison, same-unit discordant-pair evaluation, fresh-generation confirmation, and provenance-based executable audits, 
so that both the model's program conjecture and the researcher's claim that ``feedback content works'' are made falsifiable. 
Across six HumanEval+ and MBPP+ benchmark$\times$model cells with three 0.5B--1.5B frozen code models, 290 \emph{dead} task--cell 
units were evaluated, where no candidate in a cached best-of-$N$ pool ($N=8$) passed the public tier. The main run produced $7{,}000$ 
fresh seeded generations, and a preregistered diagnostic follow-up produced $1{,}400$ additional generations. True unlock required 
passing both the public and prompt-hidden EvalPlus tiers. Blind resampling exceeded bare-code retry by $+18$ net 
unlocks (discordant $25/7$, Holm-adjusted $p=0.0021$). Code-plus-facts produced $+18$ net recovery relative to 
bare code (discordant $21/3$, Holm-adjusted $p=0.00042$), and a $+15$ advantage over the same-code generic-bullet placebo supported the 
contribution of executed fact content (Holm-adjusted $p=0.0041$). In the diagnostic follow-up, an instruction-only effect 
was not distinguishable ($+3$, $p=0.36$), which is not evidence of absence. The remaining difference was associated with 
code-bearing context, although pure code causation was not claimed because code content remained confounded with prompt length. 
Code-plus-facts and blind resampling tied at the pooled level with $26$ unlocks each, which is not evidence of equivalence, 
non-inferiority, non-superiority, or unit-level identical behavior, but only zero observed pooled difference in the tested sample. 
The content-attribution contrast changed direction in a small disjoint sensitivity sample, and surface-form explanations were not 
fully ruled out. Six prospective external-controller follow-ups on the search-bound stratum further supported this boundary: 
the two strongest content channels, own near-miss code and runtime execution trace, tied with content-free shape placebo. 
The findings are limited to observed unlock outcomes and make no direct claim about latent sampling capacity, 
model distribution, or general program-synthesis ability. Taken together, falsification in the tested frozen-small-model 
regime was valuable not as Popperian vocabulary, verbal self-critique, or re-exposure to failing code, but as comparison 
with externally produced, oracle-relative, executable counterexamples. This execution-grounded criticism was observed as 
a limited signal repairing harmful code-bearing conditioning, not as a content lever exceeding blind resampling.
The transferable contribution is therefore a methodology for making self-repair feedback claims falsifiable through content-vs-form, 
discovery-vs-confirmation, and audit-vs-statistical-verdict distinctions.
\end{abstract}

\section{Introduction}
\label{sec:intro}

Small \emph{frozen code models} now appear regularly in laptop, local workstation,
and edge-device deployment settings where retraining is not feasible. In such settings, the
model is often expected to generate not a single final program, but multiple candidate programs
under a fixed generation budget. When a candidate program fails the tests, generation is repeated,
and the operational work unit is therefore formed not as a single completion but as a retry loop
\citep{wang2024software,zubair2025repair}. Within this retry loop, adding the failed program back
to the prompt, showing the evidence obtained during execution, and asking the model to produce a
corrected solution have been adopted as natural debugging practices. However, the intuitive appeal
of this practice often leads three distinct assumptions to be accepted jointly: it is assumed that
showing the failing code again is useful, that execution evidence carries an additional signal, and
that feedback-assisted retry is more effective than fresh sampling performed under the same
output-generation budget.

This problem can be formulated not only as a technical prompt-design issue, but also as a
Popperian \emph{conjectures-and-refutations} process
\citep{popper1959logic,popper1963conjectures}. The generated program can be treated as a
\emph{conjecture} whose correctness is only tentatively accepted. The violation observed in the
executed test can then be interpreted as an executable \emph{counterexample}, produced against a
specific behavioral consequence of that conjecture and relative to the test oracle. From this
perspective, the value of self-repair feedback should arise not from re-exposing the model to its
own failed output, but from opening that output to external, execution-grounded, and auditable
criticism. The central question is therefore not whether feedback is useful in general. The sharper
question is which form of refutation carries operational signal: exposure to failing code, executed
counterevidence, generic retry instruction, or blind resampling under the same budget.

The expectation that feedback may be useful is supported primarily by findings reported on stronger
models or on models specifically trained to use feedback. Iterative self-refinement has been reported
to provide improvement across different task families \citep{madaan2023selfrefine,shinn2023reflexion},
and execution-grounded debugging loops have been shown to improve performance when test results and
traces are incorporated back into the prompt \citep{chen2024selfdebug,zhong2024ldb}. It has also
been shown that 33B open models trained for multi-turn execution feedback can reach high pass rates
on HumanEval and MBPP \citep{zheng2024opencodeinterpreter}. For this reason, feedback-driven repair
is treated as one of the central workflows in large language model (LLM)-based software testing and
program-repair reviews \citep{wang2024software,zubair2025repair,takallou2026llms}.

By contrast, it has been shown that intrinsic self-correction can reduce reasoning accuracy when no
external signal is available \citep{huang2024cannot,valmeekam2023self}, that self-critique without a
sound checker can degrade symbolic reasoning \citep{stechly2025selfverification}, and that models
cannot reliably verify their own code against natural-language specifications
\citep{jin2025selfcritiquefail}. When the cost of feedback turns is taken into account,
self-repair gains have also been reported to be dominated by additional independent samples
\citep{olausson2024selfrepair}. This limitation appears especially pronounced for code models below
a few billion parameters. Reliable self-refinement is often obtained only when retraining is performed
with the relevant objective \citep{ding2024cycle,cho2025cocos,jiang2024ledex}. It has also been
reported that runtime feedback is ``not always helpful'' for small open models
\citep{tian2024debugbench}, that the generation--verification gap remains non-positive at the 0.5B
scale \citep{song2025mindthegap}, and that self-generated reasoning can reduce pass rates for
lightweight models \citep{yang2024cotton}.

In this regime, where retraining is excluded, the true competitor of feedback-assisted repair is not
another prompting recipe but the use of the same output-generation budget for fresh samples. Under
repeated sampling, coverage has been shown to scale across broad budget ranges
\citep{brown2024monkeys}, and ``more samples'' has been used as a fundamental baseline for adaptive
methods in compute-optimal test-time strategies \citep{snell2024testtime,damani2024hard}. More
elaborate verification strategies have been reported to require additional compute even to match
plain repeated sampling \citep{singhi2025solve}, while multi-turn self-repair can fall below
independent sampling under equal output budget \citep{zheng2025multiturn}. The decisive operational
question is therefore not whether a revision is better than the previous completion. It must be
tested directly whether the same generation budget should be allocated to retry with failing context
or to blind resampling after discarding that context.

This study is positioned as the third stage of a falsification-centered measurement program in which
this question has been progressively narrowed. In the first stage of the program, Popperian
code-generation skill was decomposed through a labels-only scaffold, a length-matched placebo, and an
execution oracle \citep{iscan2026scaffold}. It was shown that full Popperian procedural content did
not provide a separable execution-correctness contribution over the labels-only scaffold, and that the
gain observed in the small model was associated more with scaffold structure than with Popperian
vocabulary. The same model's self-selection using a Popperian rubric also did not exceed random
selection. In the second stage of the program, falsification was moved from prompt content to post-hoc
candidate processing \citep{iscan2026selection}. Twenty-six semantic post-hoc operators, including
selection, verification, repair, elimination, sound vetoes, and generation conditioning, are evaluated
against a Best-of-$N$ baseline under a deterministic execution oracle and a leakage-free matched-compute
protocol. No semantic output-space operator exceeded held-out accuracy over Best-of-$N$ in the tested
cells. Thus, Popperian vocabulary and model-internal post-hoc falsification are found not to carry a
separable signal, and the remaining question is directed toward whether externally executed
counterevidence can carry signal within iterative regeneration.

Four identification problems in the literature prevent this question from being answered directly.
First, positive studies often present failing code, execution evidence, and repair instruction inside
a single feedback bundle. Therefore, the component from which the reported gain arises cannot be
determined. Second, negative studies often remove the bundle entirely. As a result, it cannot be
disambiguated whether the observed harm arises from the revision act, failing code, execution evidence,
or the surface form of additional context. Third, when feedback-assisted regeneration and blind
resampling are not compared under the same output-generation budget, the practitioner's choice is not
directly tested. Fourth, aggregate pass rates over tasks that can already be solved by ordinary sampling
provide only limited information about the zero-pass-in-pool regime in which the retry loop is most
needed.

These identification problems are compounded when the independent effects of prompt surface
form are taken into account. In large-scale software-engineering evaluations, the feedback packet is
often treated as a single intervention \citep{zheng2025understanding}. Although prompt-ingredient
sweeps have begun to separate certain components, the model's own failing code is often not evaluated
as an independent arm, and same-unit outcomes are not paired through discordant-pair tests
\citep{khojah2025promptprogramming}. It is also known that prompt formatting and arbitrary prompt
architecture can affect outputs independently of task content
\citep{sclar2024formatting,brucks2025prompt}. Therefore, an ordinary no-feedback control is not
sufficient to separate task-relevant fact content from the effects of additional text presence,
template, and bullet structure.

For all these reasons, the placebo-controlled disambiguation protocol of the program is transferred
in this study to the axis of self-repair feedback content, and a five-arm controlled decomposition is
presented. Bare failing code (\armP), blind resampling (\armB), executed facts only (\armTC), code plus
facts (\armCF), and code plus generic placebo bullets (\armGR) are compared on the same task--model
units and under the same output-generation budget. Through the mirror chain, source code, failure
instruction, template, and bullet count are held fixed between \armCF{} and \armGR{}, while only
task-relevant fact content is changed. This comparison constitutes the most tightly controlled content
contrast in the design. By contrast, exact token length, lexical distribution, numeric-literal density,
literal input/output (I/O) syntax, and information density are not held fixed. The \armB{}--\armP{}
and \armCF{}--\armP{} comparisons are defined as packet-level effects. Since these arms follow
independent trajectories, they are not interpreted as same-code minimal pairs. The independent role of
failure instruction is additionally tested through a preregistered sixth arm.

The primary estimand is defined on \emph{dead} task--cell units for which no candidate in a cached
best-of-$N$ pool ($N=8$) passed the public tier. The term ``dead'' does not mean that no correct program
exists in the model distribution. It denotes only zero-pass status with respect to the finite cached
pool. Across six benchmark$\times$model cells constructed from HumanEval+ and MBPP+ with three
0.5B--1.5B frozen code models, 290 dead units are evaluated. The same output-generation budget is
allocated to each arm for each unit. Because input-token compute is not equalized, the term
\emph{matched output-generation budget} is used throughout the study. Correctness is determined by
execution on public EvalPlus tests and prompt-hidden EvalPlus tests that are not shown in the prompt
\citep{liu2023evalplus}. This design controls the advantage of producing more output samples and
allows the content, form, and exposure components of the feedback packet to be evaluated within the
same operational decision problem.

The measurement logic of the study is not limited to arm decomposition. Discovery and confirmation
are separated. Unplanned contrasts observed in the preceding preregistered stage are not used as
confirmatory evidence, but are treated only as discovery signals that generate new hypotheses. Each
effect is promoted only once to a new preregistered family, is re-measured on the same fixed units
with fresh seeds and fresh generations, and discovery outputs are not included in the confirmation
analysis \citep{popper1959logic,popper1963conjectures,mayospanos2006severe,hofman2023prereg}. Thus,
fresh-output confirmation is provided, but no independent-task replication claim is made. This
distinction is built into the measurement design to prevent exploratory evidence from being mixed
with confirmatory evidence.

This procedure also establishes Popperian falsification not only as an object-level test of programs
but also as a reflexive evaluation architecture. At the object level, generated programs are treated
as conjectures, and executed test violations are treated as counterexamples. At the meta-methodological
level, the study's own hypotheses, contrasts, and evaluation pipeline are kept open to criticism
through preregistration, placebo controls, fresh-generation confirmation, and executable audit
invariants. When an audit failure occurs, the original verdict is not deleted and the amendment is
reported separately, in order to prevent negative evidence from being neutralized by post hoc
reinterpretation. Thus, falsification is operated not only as a test directed at model outputs, but
also as a critical-control discipline to which the research system that produces claims about those
outputs is itself subjected.

No new repair method is proposed in this study. Instead, a component-level measurement methodology is
presented for measuring in what form the falsification signal is carried within the self-repair
feedback packet. The originality of this methodology is established through the decomposition of the
feedback packet into its components, the testing of surface-form explanations through content-free
placebo controls, the comparison against blind resampling under a matched output-generation budget,
the use of same-unit discordant-pair evaluation, and the separation of statistical verdict from audit
verdict. In this way, claims about self-repair feedback are made measurable not only through aggregate
pass rates, but also through the layers of content, form, sampling budget, confirmation status, and
executable provenance.

In this framework, the study makes five original contributions. The first is philosophical and
conceptual: self-repair feedback is reformulated not merely as an iterative retry practice but as a
measurable falsification problem grounded in the Popperian conjecture--counterexample relation, and by
treating generated programs as conjectures and executed test violations as oracle-relative executable
counterexamples, the distinction between exposure and criticism is made explicit, and falsification is
applied not only to object-level program outputs but also to the researcher's claims about feedback
content. The second is a measurement and evaluation contribution: a five-arm controlled decomposition
is developed to separate the components inside the feedback packet, in which bare failing code, blind
resampling, executed facts, code-plus-facts, and shape-matched placebo arms are compared on the same
task--model units and under the same output-generation budget, so that \emph{packet decomposition},
\emph{placebo mirroring}, a matched-budget baseline, same-unit discordant-pair evaluation,
fresh-generation confirmation, and provenance-based executable audits are combined within a single
reflexive measurement instrument.

The third is an identification contribution: failing code, execution evidence, and retry instruction,
which are often presented jointly in prior work, are decomposed, so that the same-code mirror contrast
separates executed fact content from generic bullet-shaped companion text and packet-level contrasts
evaluate bare-code retry, code-plus-facts retry, and blind resampling within the same operational
decision problem, leaving the measured recovery less confounded with exposure, surface form,
instruction, or an additional sampling opportunity. The fourth concerns statistical rigor and
confirmation: discovery outputs are not mixed with confirmation evidence, promoted hypotheses are
retested with fresh seeds and fresh generations, and exact McNemar tests, Holm correction, same-unit
paired evaluation, and a preregistered family structure keep claims bounded by observed contrasts,
while non-significant results, pooled ties, and descriptive guards are not converted into equivalence
or no-effect claims. The fifth is a reproducibility and audit contribution: the evaluation pipeline is
made auditable through immutable unit keys, seed-namespace isolation, frozen prompt rendering, mirror
SHA equality, forbidden-token checks, executable audit invariants, forensic audit trails, and dated
amendments, and statistical verdicts are kept separate from audit verdicts, so that not only model
outputs but also the evaluation system that produces claims about those outputs is left open to
falsification.

The claims are limited to 0.5B--1.5B frozen code models, Python HumanEval+/MBPP+ tasks, the
zero-pass-in-pool regime, a single prompt scaffold, and a matched output-generation budget. No direct
inference is made about stronger models, non-dead tasks, other programming languages, or
repository-level coding settings. The comparison between code-plus-facts and blind resampling is
interpreted outside the superiority family and is not used to produce an equivalence or
non-inferiority claim. The content-attribution interpretation is limited to the dimensions held fixed
inside the mirror chain and does not assume equality of exact token length or lexical distribution.

The remainder of the paper is organized as follows. The measurement instrument, population definition,
regeneration arms, mirror chain, preregistered statistical family, generation protocol, and audit
invariants are presented in \S\ref{sec:method}. Primary effects, sixth-arm diagnostic decomposition,
content-attribution analyses, sensitivity checks, and audit outcomes are reported in
\S\ref{sec:results}. Mechanism interpretations, prior-work comparison, practical implications, and
validity boundaries are discussed in \S\ref{sec:discussion}. The prospective external-controller
follow-ups conducted after the primary analysis are given in \S\ref{sec:controllers}. Finally, the
bounded conclusion and transferable methodological contribution of the study are summarized in
\S\ref{sec:conclusion}.

\section{Methods}
\label{sec:method}

The method used in this study was structured as the third stage of a
falsification-centered measurement program applied to feedback content. The
placebo-controlled disambiguation protocol and the leakage-free, matched-budget execution
protocol developed in the previous stages of the program were applied here to the components
of an iterative self-repair packet \citep{iscan2026scaffold,iscan2026selection}. At the object
level, the generated program is treated as a \emph{conjecture}. The violation produced by
public test execution is interpreted as an execution-grounded and oracle-relative
\emph{counterexample} to that conjecture
\citep{popper1959logic,popper1963conjectures}. At the meta-methodological level, the strength
of the executed counterexample is taken to depend on assumptions about the test oracle,
sandbox, and harness. Because of this Duhem--Quine boundary, the arm contrasts, statistical
claims, and the evaluation pipeline that produced those claims are kept open to critical
testing through preregistration, fresh-generation confirmation, and executable audit
invariants.

\begin{figure}[t]
\centering
\includegraphics[width=\linewidth]{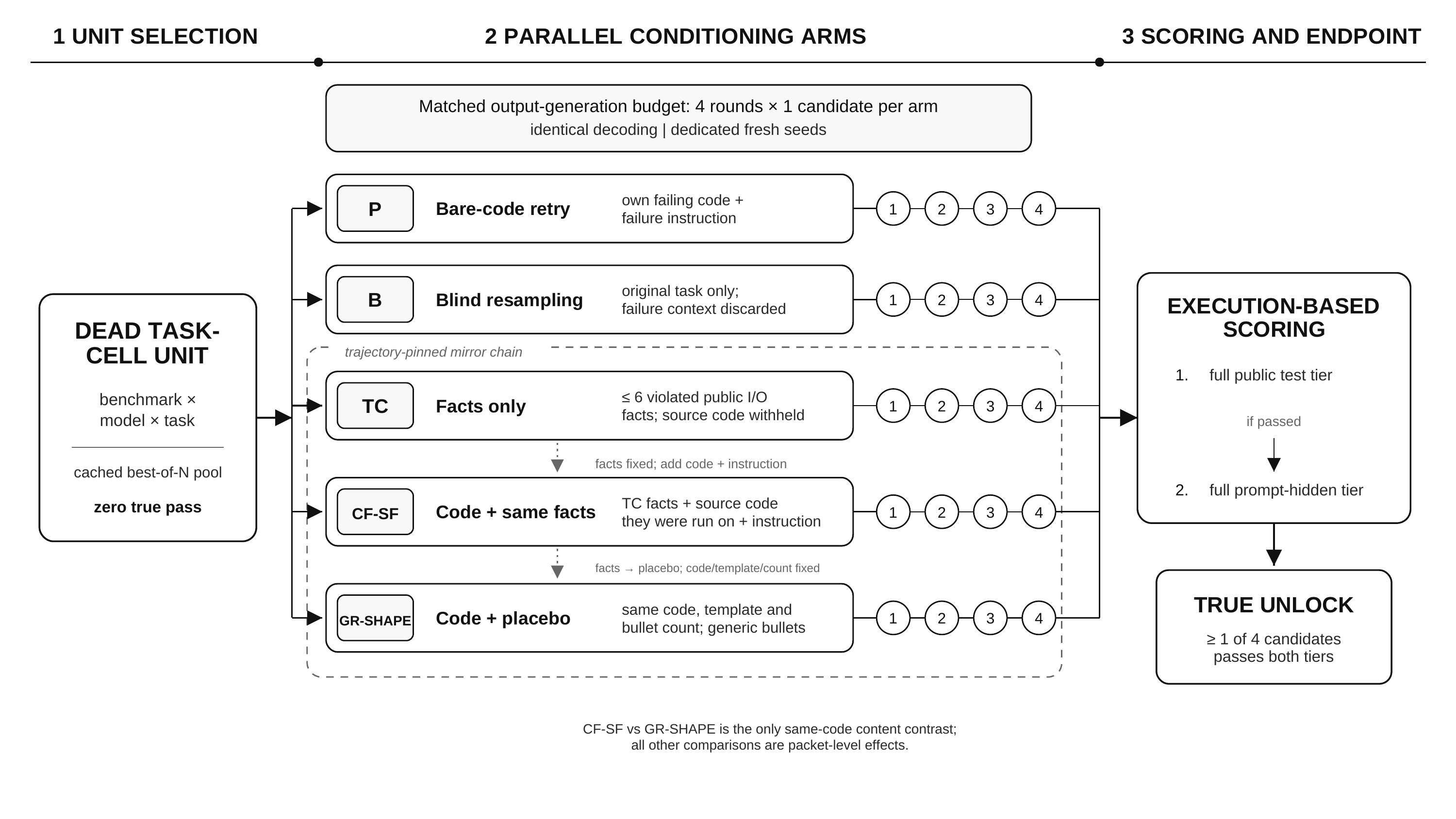}
\caption{Falsification-centered measurement instrument. Each dead unit for which no candidate
in the cached pool passed the public tier was evaluated under five regeneration arms with the
same output-generation budget. Four rounds were run for each arm, one candidate was generated
in each round, and a fresh seed from a dedicated namespace was used. The mirror chain links the
\armCF{} packet to the \armTC{} trajectory and the \armGR{} packet to the \armCF{} trajectory.
Only in the \armCF{}--\armGR{} comparison were source code, failure instruction, template, and
bullet count held fixed while task-relevant fact content was changed. The endpoint is true
unlock. At least one candidate generated under the same arm must pass both the public and
prompt-hidden tiers.}
\label{fig:design}
\end{figure}

A five-arm controlled decomposition was used to measure the component-level effects of the
self-repair feedback packet. Each dead task--cell unit was evaluated under five parallel
regeneration arms with a matched output-generation budget. The outcome was defined by whether
at least one candidate generated under the same arm passed both the public and prompt-hidden
test tiers. The measurement instrument is summarized in Figure~\ref{fig:design}. This section
describes the population and endpoint, arm definitions, mirror chain, preregistered statistical
family, generation protocol, audit invariants, and discovery-to-confirmation procedure in
sequence. Procedure~\ref{alg:procedure} is given not as a new repair algorithm, but as the
executable evaluation procedure of the measurement instrument.

\subsection{Measurement instrument}
\label{sec:design}

\subsubsection{Population, endpoint, and unit of analysis}
\label{sec:regime}

\begin{definition}[Dead task--cell unit]
\label{def:dead}
A \emph{cell} is defined as a benchmark--model pair. A \emph{task--cell unit} is the
benchmark task evaluated under a specific cell and is immutably keyed by
$\langle\textit{benchmark}\,|\,\textit{model}\,|\,\textit{task}\rangle$.
A unit is classified as \emph{dead} when none of the candidates in the cached best-of-$N$
pool for that unit ($N=8$, fixed by the \texttt{K\_POOL} code constant) passes the public
test tier under standard and robust extraction. The per-task candidate cache files are an
uncommitted upstream artifact. This definition is made with respect to the finite cached pool
and does not imply that no correct program exists in the model distribution.

\end{definition}

\begin{definition}[True unlock]
\label{def:unlock}
A dead unit is counted as \emph{unlocked} by a given arm only if at least one regenerated
candidate from that arm passes the full public tier \emph{and} the full prompt-hidden tier.
Prompt-hidden test cases are not shown to any arm. However, because the dead population is
defined only by public-tier pass/fail outcomes in the cached pool, and because the
prompt-hidden tier defines the final true-unlock scoring endpoint, this tier is interpreted
not as an untouched test set but as a prompt-hidden evaluation tier.
\end{definition}

The population was constructed from 512 dead units obtained by crossing the HumanEval+ and
MBPP+ benchmarks \citep{chen2021humaneval,austin2021mbpp,liu2023evalplus} with three frozen
models served through Ollama. The models were fixed as
\texttt{qwen2.5-coder:1.5b}, \texttt{qwen2.5-coder:0.5b-instruct} (fp16), and
\texttt{deepseek-coder:1.3b} \citep{hui2024qwen25coder,guo2024deepseekcoder}. The seeded
and capped fixed sample used in the predecessor experiment was inherited unchanged. After the
inherited pre-run token exclusion, the primary confirmation sample was fixed at $n=290$.
Thus, the promoted hypotheses are retested on the same task--cell sample in which the
discovery effects were produced, but with entirely fresh generations.

A separate disjoint sample consisting of 60 dead units that had not previously been included
in the primary sample was constructed for external-direction sensitivity. This sample contains
15 HumanEval+/deepseek units, 30 MBPP+/deepseek units, and 15 MBPP+/qwen-0.5b units.

The per-cell sample flow is summarized in Table~\ref{tab:sampleflow}. In each cell, dead units
were first capped at $\mathrm{cap}=60$ per cell, and the pre-run token exclusion was then
applied. The only exclusion was the \texttt{humanevalplus\_\_115} unit in the
HumanEval+/\texttt{deepseek-coder:1.3b} cell. Therefore, the capped 60 units in that cell were
reduced to 59 primary units, yielding a total primary sample of $290$ and a capped total of
$291$. The disjoint sample was drawn from the unselected units of each cell. A 32-bit seed
derived by $\mathrm{sha256}$ from the payload
\texttt{antidote-disjoint|benchmark|model} and \texttt{random.Random.sample} were used for
this selection. The cached best-of-$N$ pool was fixed at $N=8$ (\texttt{K\_POOL} code
constant). Per-task candidate cache files were retained as an uncommitted upstream artifact.

The provenance of the public and prompt-hidden tiers is given in Table~\ref{tab:tierprov}.
The public tier consists of EvalPlus base tests. This tier is used by the evaluation harness,
but it is not rendered verbatim in any arm prompt. In the \armTC{} and \armCF{} arms, only up
to six public assertions violated by the incumbent are shown as literal input/output (I/O)
facts. No test-derived content is provided to the \armB{} and \armP{} arms. The \armGR{} arm
is instead given count-matched generic placebo bullets. The prompt-hidden tier consists of
EvalPlus plus tests and is deterministically capped at the first 300 tests in EvalPlus source
order. This tier is not shown to any arm. Dead-status selection is performed only on the
public tier, whereas the prompt-hidden tier is used only for final true-unlock scoring.
Therefore, the prompt-hidden tier is referred to as the prompt-hidden evaluation tier that
defines the final scoring endpoint and is not presented as an untouched held-out test set.

\begin{table}[t]
\centering\footnotesize\setlength{\tabcolsep}{4pt}
\caption{Per benchmark$\times$model cell sample flow. Dead units were first capped at
$\mathrm{cap}=60$ per cell, and pre-run token exclusion was then applied.
Primary $=$ capped $-$ exclusions. The capped total is $291$, and the single exclusion
(\texttt{humanevalplus\_\_115}) reduces the primary sample to $290$. Initial counts were
computed from the deterministic EvalPlus loader (MBPP+ $378$ nominal minus $13$ loader
exclusions $=365$, HumanEval+ $164$). All counts were recomputed from committed ANTIDOTE
prerun manifests with \texttt{scripts/reconstruct\_sample\_selection.py} and reproduced
under the seeded selection rule.}
\label{tab:sampleflow}
\fitwidth{\input{tables/sample_flow}}
\end{table}

\begin{table}[t]
\centering
\caption{Test-tier provenance. The public tier ($\Ipub$, EvalPlus base input suite) is
available to the harness but is not shown verbatim in any arm. It appears only in the
\armTC{}/\armCF{} arms as $\le 6$ violated input/output (I/O) facts. The prompt-hidden tier
($\Ihid$, EvalPlus plus input suite, first $300$ in source order) is not shown to any arm.
Dead-status selection depends only on public-tier outcomes. The prompt-hidden tier is not used
for selection and is evaluated only during final true-unlock scoring. Therefore, it is not an
untouched held-out set. All tiers use EvalPlus 0.3.1, pinned in \texttt{uv.lock}.}
\label{tab:tierprov}
\fitwidth{\input{tables/test_tier_provenance}}
\end{table}

\begin{table}[t]
\centering\footnotesize
\caption{Prompt-hidden test-count summary. The deterministic first-$300$ cap binds only on
HumanEval+ ($126$ tasks truncated, pre-cap median $972$, max $1000$). On MBPP+ the cap never
binds (max $147$). The cap binds the prompt-hidden tier only. Dead-status selection uses
public-tier outcomes only, so the cap affects final true-unlock scoring rather than dead
selection.}
\label{tab:testcount}
\fitwidth{\input{tables/test_count_summary}}
\end{table}

Units are selected by seeded hash from the eligible pool remaining outside the fixed sample
(\S\ref{sec:robustness}). The disjoint sample is not included in the primary statistical
family and is used only for directional sensitivity.

Cell-level dead-task counts, model digests, and round-0 prompts are recorded as SHA-pinned
version-controlled manifests before any generation is initiated. This prevents sample identity
and prompt rendering from being changed after the run.

The matched budget was operationalized as an equal output-sample count per unit. Four
regeneration rounds were run for each arm, one candidate was produced in each round, and
identical decoding parameters were used. Thus, four output samples were allocated to each arm
for each unit. Input-token compute was not equalized. Code-bearing prompts reached approximately
1{,}290--1{,}480 characters depending on the arm, compared with a mean length of approximately
406 characters for \armB{} prompts. Therefore, strict end-to-end floating-point operation (FLOP)
matching was not provided. Instead, a \emph{matched output-generation budget} was provided.
Outcome differences cannot be attributed to generating more candidates. Prompt length and
prefill cost remain residual design factors
\citep{olausson2024selfrepair,brown2024monkeys,zheng2025multiturn}.

\subsubsection{Regeneration arms and mirror chain}
\label{sec:arms}

The five arms shown in Figure~\ref{fig:design} are defined in Table~\ref{tab:arms}. In the
\armP{} condition, the model was shown its own round-specific best-so-far failing code and was
asked to produce a corrected solution. In each round, the arm-specific incumbent was selected
from the union of the cached best-of-$N$ pool and the candidates generated in the previous
rounds of that arm. Selection was based on the highest public-assert pass count, namely the
number of public assertions passed. When pass counts were tied, ties were resolved
deterministically in favor of the most recently generated candidate. Thus, under a tie, the
latest round output rather than a pool entry was accepted as the incumbent. In the \armB{}
condition, no failure context other than the original task was provided, and each round was
run as an independent blind resample. These two arms represent the bare-code retry practice
and the discard-the-context baseline, respectively.

In the \armTC{} condition, only execution-grounded counterevidence was presented. Public
assertions violated by the incumbent solution in the corresponding trajectory were shown as at
most six literal I/O facts through a frozen no-ellipsis renderer. These facts were used as
observable counterexamples that refuted a specific behavioral consequence of the candidate's
correctness conjecture in an oracle-relative manner. The source code on which the facts had
been executed was withheld from the prompt. If the incumbent violates no public assertion, that
is, if all public tests are passed and failure occurs only on the prompt-hidden tier, the
fact block is rendered as an empty block containing only the header line and no bullets. This
empty-violation behavior is shared by the \armTC{}, \armCF{}, and \armGR{} arms. Since the
\armGR{} placebo bullet count mirrors the true fact count in \armTC{}, zero true facts
correspond to zero placebo bullets. In this case, the CONFIRM-local renderer inserts no
degenerate fallback sentence.

The three code-bearing arms, \armP{}, \armCF{}, and \armGR{}, were terminated with the
following failure instruction: ``Your solution failed the tests. Write a corrected, different
solution.'' The two code-free arms, \armB{} and \armTC{}, did not include this instruction.
Therefore, in the main five-arm design, the failure instruction remained coupled with
code-bearing context. This coupling was tested by a separate preregistered sixth-arm
follow-up.

\begin{table}[t]
\caption{Five regeneration arms. Four output generations were used per unit for each arm. In
the \armCF{}--\armGR{} mirror pair, source code, failure instruction, template, and bullet
count were held fixed, and only task-relevant fact content was changed. The
\armTC{}--\armCF{} transition adds source code and the failure instruction together.
\armP{} comparisons involve independent arm trajectories and are therefore not interpreted as
same-code minimal pairs.}
\label{tab:arms}
\centering\footnotesize
\fitwidth{\begin{tabular}{llccc}
\toprule
Arm & Conditioning packet & Own code & Executed facts & Companion text \\
\midrule
\armP{}  & bare own failing code (incumbent)             & \checkmark & --- & --- \\
\armB{}  & none (blind resample of the task)             & --- & --- & --- \\
\armTC{} & violated public I/O facts only ($\le 6$)      & --- & \checkmark & --- \\
\armCF{} & \armTC's facts + the code they were run on    & \checkmark & \checkmark & --- \\
\armGR{} & \armCF's code + shape-matched generic bullets & \checkmark & --- & \checkmark \\
\bottomrule
\end{tabular}}
\end{table}

The \armCF{} and \armGR{} arms were constructed as mirror conditions. In the \armCF{}
condition, the source code from which the \armTC{} fact block had been produced was shown in
full, and the failure instruction was added because of the code-bearing scaffold. The rendered
fact block was copied \emph{verbatim} from the \armTC{} packet, and equality was verified by
per-round SHA-256 audit. Thus, the \armTC{}--\armCF{} transition added source code and the
failure instruction together while evidence was held fixed. This comparison is not interpreted
as a code-only effect.

The \armGR{} condition was constructed as a shape-matched placebo. In this arm, the \armCF{}
source-code block was preserved exactly and verified by per-round SHA-256 equality. The
failure instruction, template, and bullet count were also held fixed. Executed facts were
replaced with bullets selected from a frozen list of generic requirement statements. Placebo
bullets were constructed so that they contained no function name, assert literal, or task
identifier, and an executable forbidden-token audit was applied to all rendered prompts.

Thus, in the \armCF{}--\armGR{} contrast, source code, failure instruction, template, and
bullet count are held fixed while task-relevant fact content is changed. This comparison is
designed to separate the effect of executed counterevidence from the presence and shape effects
of generic companion text. Nevertheless, because exact token length and lexical distribution
are not fully equalized, it is not claimed that all surface-form differences are removed.

It has been shown that meaning-preserving changes in prompt form can produce large differences
in accuracy \citep{sclar2024formatting}, and that arbitrary prompt architecture can bias
outputs independently of task content \citep{brucks2025prompt}. Therefore, the placebo arm is
used as the load-bearing control for feedback-content attribution.

To test the code--instruction coupling in the main design, a preregistered sixth arm named
\armBI{} (blind plus instruction) was run as a sequential follow-up. \armBI{} was constructed
by adding only the failure instruction used in the code-bearing arms to the \armB{} scaffold.
No code, facts, or bullets were added. Four rounds were run on the same fixed and disjoint
units, a separate seed namespace was used, and outcomes were paired at the unit level with the
frozen \armB{} and \armP{} outcomes. Thus, the instruction component and the code-bearing-context
component were measured through separate contrasts. Code content and the intrinsic prompt
length introduced by code are not decomposed in this follow-up.

\begin{algorithm}[t]
\caption{Five-arm evaluation procedure. The procedure is not a repair algorithm, but an
evaluation instrument that measures the arm-level true-unlock outcome for each dead unit under
a matched output-generation budget.}
\label{alg:procedure}
\begin{algorithmic}[1]
\Require Dead unit set $\mathcal{U}$, arms
$\mathcal{A}=\{\armP,\armB,\armTC,\armCF,\armGR\}$, rounds $R=4$,
frozen packet renderer, frozen decoding parameters
\Ensure Per-unit and per-arm outcome $\mathrm{unlock}(u,a)$
\ForAll{dead units $u \in \mathcal{U}$}
  \For{$r=1$ to $R$}
    \State retrieve the frozen round-$r$ trajectory state for each arm
    \State render $\pi_{u,\armP,r}$ from the \armP{} failing incumbent
    \State render $\pi_{u,\armB,r}$ from the original task only
    \State render $\pi_{u,\armTC,r}$ from at most six violated public I/O facts
    \State render $\pi_{u,\armCF,r}$ by copying the \armTC{} fact block and
    adding the source code on which those facts were executed
    \State render $\pi_{u,\armGR,r}$ by copying the \armCF{} code block and
    replacing the fact block with count-matched placebo bullets
    \State verify mirror SHA equality and run the forbidden-token audit
    \ForAll{arms $a \in \mathcal{A}$}
      \State derive fresh seed $s_{u,a,r}$ from the dedicated namespace
      \State $\hat{c}_{u,a,r}\gets
      \mathrm{decode}(\pi_{u,a,r},s_{u,a,r})$
      \State execute $\hat{c}_{u,a,r}$ on the public tier
      \If{$\hat{c}_{u,a,r}$ passes the public tier}
        \State execute $\hat{c}_{u,a,r}$ on the prompt-hidden tier
      \EndIf
      \State record the candidate, scores and frozen transition state
    \EndFor
  \EndFor
  \ForAll{arms $a \in \mathcal{A}$}
    \State $\mathrm{unlock}(u,a)\gets
    \mathbf{1}\!\left[\exists r\le R:
    \hat{c}_{u,a,r}\text{ passes both tiers}\right]$
  \EndFor
\EndFor
\end{algorithmic}
\end{algorithm}

\subsection{Statistical analysis, generation protocol, and disclosure}
\label{sec:analysis}

\subsubsection{Preregistered statistical family}
\label{sec:stats}

The frozen statistical family consisted of three one-sided claims evaluated on pooled
discordant pairs:

\begin{itemize}[leftmargin=2em]
  \item \Cone{} (co-primary, \emph{trap} hypothesis): $\armB>\armP$.
  \item \Ctwo{} (co-primary, \emph{antidote} hypothesis): $\armCF>\armP$.
  \item \Cthree{} (confirmatory secondary, \emph{content attribution}):
        $\armCF>\armGR$.
\end{itemize}

Each claim was evaluated using an exact one-sided McNemar test
\citep{mcnemar1947note,fagerland2013mcnemar}. For the favored-only discordant count $\bzo$
and the reverse-only count $\bten$, the exact probability was computed as

\begin{equation}\label{eq:mcnemar}
p_{\mathrm{exact}}=
\sum_{k=\bzo}^{\bzo+\bten}
\binom{\bzo+\bten}{k}
\left(\tfrac{1}{2}\right)^{\bzo+\bten}.
\end{equation}

Here, $\bzo$ denotes units unlocked by the favored arm but not unlocked by the comparison arm.
$\bten$ denotes units unlocked only by the comparison arm. Concordant pairs do not contribute
to the test statistic. Same-unit paired-binary evaluation allows the unique unlock pattern of
each contrast to be measured instead of an aggregate pass-rate difference.

Familywise error for the three $p$-values was controlled by Holm correction
\citep{holm1979simple}. For ordered values $p_{(1)}\le\dots\le p_{(m)}$, the first threshold
violation is defined as

\begin{equation}\label{eq:holm}
k=\min\Big\{\,i:
p_{(i)}>\tfrac{0.05}{m-i+1}\,\Big\}.
\end{equation}

$H_{(1)},\dots,H_{(k-1)}$ are rejected, and all hypotheses are rejected if no violation is
found. Here, $m=3$ and the familywise significance level was fixed at $0.05$.

Under the frozen statistical rule, a positive family verdict required both co-primary claims
to satisfy the Holm-adjusted $p<0.05$ condition. If only one co-primary claim passed the
threshold, the verdict was defined as NOT-CERTIFIED with a named split, and partial
certification was not applied. Passing the threshold for \Cthree{} allowed the antidote result
to be described as ``content-attributed.'' Failure of \Cthree{} alone did not block the
co-primary statistical verdict.

The term ``certified'' was used only as the operational label of the pre-committed statistical
decision rule. Preregistration was internal and version-control-timestamped. No registered
report using OSF, AsPredicted, or in-principle acceptance was used. The statistical verdict and
audit verdict were kept separate. If the frozen audit rule produced a failure, the original
audit status was preserved as AUDIT-FAILURE even if the statistical thresholds had been passed.
The amended verdict computed after the dated amendment was reported separately, and the
original verdict was not deleted.

Each contrast was paired through immutable task--cell keys. The evaluator executable asserted
that the pooled $n$, $\bzo$, and $\bten$ values were equal to the sums of the six per-cell
tables. This invariant was added after the pooling-collapse bug detected in the discovery
stage.

To make the effect direction visible beyond pooled counts, the Haldane--Anscombe-smoothed
pooled matched-pair odds ratio is reported as

\begin{equation}\label{eq:orpool}
\widehat{\mathrm{OR}}_{\mathrm{pool}}=
\frac{\sum_{s=1}^{6} b_{01,s}+0.5}
     {\sum_{s=1}^{6} b_{10,s}+0.5}.
\end{equation}

Here, $b_{01,s}$ and $b_{10,s}$ are the favored- and reverse-direction discordant counts
within stratum $s$. This quantity is not a Cochran--Mantel--Haenszel (CMH) estimator. It is
used as a pooled discordant odds ratio for descriptive effect-size summary, together with
per-cell signs and guard checks.

Two additional preregistered descriptive guards are applied to constrain claim wording to
observed effects. The \emph{material-rescue} guard requires the $\armGR-\armP$ rescue to be
less than half of the $\armCF-\armP$ rescue for content-attributed wording. The
\emph{observed deployment-harm} gate applies the condition

\begin{equation}\label{eq:harm}
\frac{\bten}{n}\le 0.02
\end{equation}

separately to each co-primary contrast. This gate is based on observed reverse-harm share. It
does not provide a confidence bound, non-inferiority proof, or deployment safety guarantee for
the population-level harm rate. Therefore, the gate is interpreted only as a preregistered
wording constraint.

Power is computed before the run as exact-binomial conditional power based on discovery-stage
discordant counts:

\begin{equation}\label{eq:power}
1-\beta=
\Pr\!\left[
p_{\mathrm{exact}}(X,D-X)
\le \alpha_{\mathrm{Holm}}
\right],
\qquad
X\sim\mathrm{Binomial}(D,\pi),
\end{equation}

where $D$ is the expected discordant-pair count, $X$ is the favorable-direction discordant
count, and $\pi$ is the favorable share. The conservative first-step Holm threshold
$\alpha_{\mathrm{Holm}}=0.05/3$ is used. Pre-run power is computed as $0.72$ for \Cone{},
using $D\approx38$ and $\pi=0.737$. For \Ctwo{}, power is computed as $0.83$, using
$D\approx36$ and $\pi=0.75$. Power drops to the $0.16$--$0.38$ range at half the discovery
effect. Therefore, it is pre-committed that a non-significant but direction-preserving result
is interpreted as ``failed to confirm'' rather than as ``refuted.''

The comparison of whether the tested arms exceeded blind resampling is not included in the
frozen superiority family. Since equality or non-superiority cannot be certified as a positive
result in a one-sided superiority test, the blind-resampling bound is reported with
descriptive discordant counts.

\subsubsection{Generation protocol and audit invariants}
\label{sec:protocol}

Generation parameters were frozen before the run, as shown in Table~\ref{tab:config}. Each
generation used 1{,}024 completion tokens, a 4{,}096-token context, one candidate per round,
and temperature $0.8$. Sampler settings were inherited unchanged from the discovery stage.

The in-run prompt-token guard capped prompts at 3{,}072 tokens. The pre-run preflight excluded
the unit whose round-0 prompt might exceed the effective 2{,}944-token bound. The primary
sample preserved the single inherited exclusion from the predecessor experiment, and no
additional unit was excluded during the current main run.

Each generation is seeded through a dedicated namespace:

\begin{equation}\label{eq:seed}
s=
\mathrm{sha256}\!\big(
\texttt{antidote|}
\langle\textit{cell}\rangle|
\langle\textit{task}\rangle|
\langle\textit{arm}\rangle|
\langle\textit{round}\rangle
\big)
\bmod 2^{32}.
\end{equation}

Bar-delimited fields keyed a single generation. The full seed payload was stored in each
record. An executable audit verified that the 7{,}000 payloads in the main five-arm run were
unique and contained no foreign namespace prefix. A separate namespace was used for the
sixth-arm follow-up.

Generated candidate code was executed inside a sandbox with a 2~GiB address-space limit. An
exclusive serving lock restricted generation requests to a single job at a time and reduced
concurrent-serving interference.

The audit layer was implemented not as an implementation detail, but as the falsification
layer of the method. During the pre-run stage, per-cell render manifests, SHA-pinned round-0
prompts, constraint-render audits, and model digests were committed to version control. During
evaluation, the following invariants were applied executably: transition completeness, exactly 5 arms
$\times$ 4 rounds $\times$ included units, mirror equality, per-round source-code and
constraint-block SHA equality, forbidden-token audit on all \armGR{} prompts, seed uniqueness and
namespace isolation, provenance-based anti-reuse checks against discovery records, and the
pooled-equals-cell-sum invariant.

The full implementation was gated before the first token was produced by a falsification suite
containing eight executable counterexamples. The suite covered mirror integrity, seed
isolation, sample identity, prompt-hidden-test isolation, and hostile-output resilience
conditions.

Under the frozen rule, any audit failure assigns the original verdict as AUDIT-FAILURE. One
audit episode was triggered during the main run and one during the sixth-arm follow-up.
Original verdicts, forensic analyses, dated amendments, and amended verdicts are reported
separately (\S\ref{sec:audit}, \S\ref{sec:deconf}). Generation outputs, unit identities, and
scores are not changed during amendments. Thus, failures of the pipeline itself are not
deleted post hoc, and the claim-producing system is also subjected to critical testing.

\begin{table}[t]
\caption{Frozen experimental configuration. Decoding parameters, arm set, sample sizes, and
seed namespace were committed to version control before the main run began. The disjoint-60
sample was used only for sensitivity analysis and was not included in the primary statistical
family.}
\label{tab:config}
\centering\small
\fitwidth{\input{tables/config}}
\end{table}

\subsubsection{Discovery and single-promotion rule}
\label{sec:discovery}

The promoted hypotheses were derived from a previously run preregistered family in the same
research program, named CONFIRM-TC. This family tested whether facts-only conditioning
(\armTC) exceeded bare code (\armP) and produced a mixed result. The first claim passed the
threshold with a $+15$ effect (Holm $p=0.0301$). The corpse-mechanism claim reversed with
$-3$ ($p=0.78$), and the family did not receive a positive verdict.

In the same run, two unplanned contrasts produced larger descriptive effects:
$\armB-\armP=+18$ ($p=0.0025$) and $\armCF-\armP=+18$ ($p=0.0020$). Of the ten possible
undirected pairwise contrasts among the five arms, five were inspected. Under a worst-case
Bonferroni correction applied to all ten contrasts, the values would remain approximately
$0.025$ and $0.020$. Nevertheless, these observations are not counted as confirmatory
evidence and are used only as screening signals that generate new hypotheses.

The promotion rule was fixed before the new run began. Each descriptive effect was allowed to
be promoted to a fresh preregistered family only once. Discovery outputs were not re-included
in the confirmation analysis. If the fresh test failed, second promotion was prohibited. This
rule was applied to prevent a failed family from being protected from refutation through post
hoc reinterpretation or repeated testing
\citep{popper1959logic,popper1963conjectures,mayospanos2006severe,hofman2023prereg}.

To keep the estimand on the same task population, task lists, cached pools, and incumbents were
preserved. Seeds and generated tokens were completely renewed. Therefore, the procedure provided
fresh-generation confirmation. It did not provide independent-unit or independent-population
replication.

\subsection{Post-primary external-controller follow-up instruments}
\label{sec:controller-methods}

In addition to the primary preregistered pipeline, six external controllers that guide the
frozen small model at inference time on the search-bound stratum are evaluated as prompt-only
follow-up instruments. These instruments are not part of the primary frozen evaluation
pipeline, are not preregistered, and are not subjected to that pipeline's audit invariants.
Their mechanisms are summarized here for method completeness, whereas their results are reported
in Table~\ref{tab:controllers}. Each controller is compared with blind resampling (\armB) and
its own content-free shape placebo under a matched output-generation budget ($R{=}4$). The
shared suffixes in the names refer to joint-embedding predictive architecture (JEPA) and
reinforcement learning (RL) components. The deployed arms are mostly deterministic fixed-rule
variants rather than learned controllers. The learned stack is deferred. Only FJR-JEPA-RL
deploys learned LinUCB selection online.

Below, the \emph{deployed} mechanism of each controller is summarized in a compact mathematical
form derived from the source code. Components that were designed but not run in this study are
marked as \emph{deferred}. Common notation is defined as follows. A \emph{unit}
$u=(b,m,\tau)$ denotes benchmark $b$, frozen model $m$, and task $\tau$. Each arm runs
$R{=}4$ rounds. The incumbent $c_{u,r}$ at round $r$ is defined as the candidate with the
highest public-pass count up to that point. For public test $j\in\{1,\dots,J_u\}$, with pass
flag $f_j\in\{0,1\}$, the cached incumbent's public-pass fraction and the search-bound stratum
are defined as

\begin{equation}\label{eq:q0sb}
q_0(u)=\frac{s_u}{J_u},
\qquad
\mathcal{S}_{\mathrm{SB}}=\bigl\{\,u:\ \tfrac{1}{3}\le q_0(u)<1\,\bigr\}.
\end{equation}

Here, $s_u=\sum_j f_j$ is the number of passed public assertions. True unlock $Y_a(u)$ is
defined as at least one candidate generated under arm $a$ passing both the public and
prompt-hidden tiers in a round (Definition~\ref{def:unlock}). \mbox{FJR-,} RIFT-, and
DCH-JEPA-RL share a falsification-severity primitive:

\begin{equation}\label{eq:sev}
\sigma(e)=T(e)\,\cdot\,\mathbf{1}[e\ \text{refutes}\ h]\,\cdot\,\frac{1-p}{1+p},
\qquad p=P(e\mid b).
\end{equation}

Here, $e$ is an executed public counterexample. $T(e)\in[0,1]$ denotes the oracle-trust value
associated with the evidence kind. $T(e)=1.0$ was used for public assertions, $T(e)=0.95$ for
metamorphic evidence, and $T(e)=0$ for model-generated evidence. $p$ is the expectedness of
the evidence under the public background. Surprising refutations ($p\to0$) receive maximum
severity. All controllers operate on public-only state and read the hidden tier only during
final true-unlock scoring.

\paragraph{FJR-JEPA-RL.}
FJR-JEPA-RL (Falsification-JEPA Reinforced Resynthesis) is used as a deployed learned
controller. On public-only context $x$, one LinUCB ridge estimator is maintained for each arm
in the action set
$\mathcal{A}=\{\armB,\textsf{CE},\textsf{MR},\textsf{SPEC},\textsf{DIV}\}$ and updated with
the observed public reward $r$:

\begin{equation}\label{eq:fjr-linucb}
A_a \leftarrow A_a + x x^\top,\qquad
b_a \leftarrow b_a + r\,x,\qquad
\hat{\theta}_a = A_a^{-1} b_a .
\end{equation}

The per-round action is selected as the argmax of an additive score. Ties are resolved with
a deterministic \texttt{sha256} tie-break:

\begin{equation}\label{eq:fjr-score}
\begin{aligned}
Q(s,a)={}&w_q\bigl[V(F(z,a))-V(z)\bigr]
+\alpha\sqrt{x^\top A_a^{-1} x}
+\eta\,\widehat{\Delta\kappa}(a)\\
&+w_N\,N(a)
-w_{\mathrm{anc}}\,\mathrm{AnchorRisk}(a)
-w_C\,c(a).
\end{aligned}
\end{equation}

Here, $F$ denotes the JEPA latent transition predictor, $V$ the public-trained value proxy,
$z$ the public-only latent state, and $s$ the policy state. $\kappa$ is the corroboration
score. $\widehat{\Delta\kappa}(a)$ defines the expected one-step corroboration gain of action
$a$ through the severity in Eq.~\eqref{eq:sev}. $N$ denotes behavioural novelty, $\alpha$ the
LinUCB exploration constant, and $w_q,\eta,w_N,w_{\mathrm{anc}},w_C$ fixed score weights. The
last two terms are anchoring-risk and compute-cost penalties. $w_{\mathrm{anc}}$ corresponds
to the \texttt{rho} field in the code. When the value head is unfit, $V\equiv 0.5$ is returned,
so the JEPA term is exactly $0$ at cold start. The hidden-label value head is offline-only and
is not used in the online run.

\paragraph{RIFT-JEPA-RL.}
RIFT-JEPA-RL (Refutation-Inversion Falsification Thought loop) is used as a deployed
fixed-rule conservative gate. For ten symbolic assumption-inversion axes, severity is
aggregated per axis from fired refuters, and a conservative threshold rule is applied:

\begin{equation}\label{eq:rift}
\rho_a=\!\!\max_{r:\,\mathrm{axis}(r)=a}\!\!\sigma(e_r),
\qquad
a^\star=\operatorname*{arg\,max}_a \rho_a,
\qquad
\text{fire } a^\star \iff \rho_{a^\star}>\delta,\ \text{otherwise } \armB .
\end{equation}

In the deployed M1 configuration, $\delta{=}0$ is used. The recipe for a fired axis is
produced by closed-form inversion of the refuted claim:
$\bar H=H\oplus\mathrm{flip}(c^\star)$. The M2 path, which extends the same gate to the form
of a pessimistic $Q^-=\hat Q-\beta\,\sigma_Q$ with learned advantage and epistemic
uncertainty, is left deferred.

\paragraph{DCH-JEPA-RL.}
DCH-JEPA-RL (Dialectical Counter-Hypothesis JEPA-RL) is defined as a designed-only and
unrealizable instrument. Falsified assumptions are designed to be converted into a
version-space bank $\mathcal{H}_t$ consisting of $K$ competing counter-hypothesis cards and
are formulated to be updated by Popperian exponential elimination:

\begin{equation}\label{eq:dch-belief}
\mu_{t+1}(i)\ \propto\ \mu_t(i)\,\exp\!\Bigl(-\lambda\!\sum_{e\in E_t}\!\ell(H_i,e)\Bigr),
\qquad
\ell(H_i,e)=T(e)\,\mathbf{1}[H_i\not\models e] .
\end{equation}

Action selection is defined by a pessimistic lower-confidence advantage computed through an
offline group-disjoint JEPA world model $F_\psi$ and hidden-transfer value head $V_\omega$:

\begin{equation}\label{eq:dch-q}
\begin{aligned}
Q^-(s,a)&=\hat Q(s,a)-\beta\,\hat\sigma_Q(s,a),\\
a_t&=\operatorname*{arg\,max}_a \mathrm{LCB}(a)\ \ \text{if}\ \ \max_a \mathrm{LCB}(a)>\delta_t,\ \ \text{otherwise } \armB .
\end{aligned}
\end{equation}

Here, $\mathrm{LCB}(a)=\widehat{\Delta}(a)-\beta\,U(a)$ is defined as the lower-confidence
bound. $\lambda$ is the Popperian elimination temperature, $\beta$ is the pessimism or
lower-confidence coefficient, and $U$ is total epistemic uncertainty. $\ell$ is the per-card
elimination penalty and is distinct from RIFT's $\rho_a$ severity aggregate. This stack is not
run. Because the required offline transition corpus is not available on disk, the controller
cannot be instantiated. Therefore, Eq.~\eqref{eq:dch-belief}--\eqref{eq:dch-q} is
classified entirely as designed-but-deferred.

\paragraph{ECK-JEPA-RL.}
ECK-JEPA-RL (Executable Constraint Kernel JEPA-RL) is used as a deployed fixed-rule kernel.
For a search-bound unit, the public assertions violated by the incumbent are accepted as an
executable constraint kernel $K$ only when they can be safely serialized into a Python
literal and parsed with \texttt{ast.parse}:

\begin{equation}\label{eq:eck}
\begin{aligned}
\mathrm{fired}(K)&=\Bigl(\exists j:\ \neg f_j\ \wedge\ \mathrm{serializeOK}(j)\Bigr)\wedge
\mathrm{astParse}\bigl(\mathrm{block}(K)\bigr),\\
a_t&=\begin{cases}\textsf{ECK}_2,&\mathrm{fired}(K)\\[2pt] \armB,&\text{otherwise.}\end{cases}
\end{aligned}
\end{equation}

Failing witnesses are ranked with the fixed 7-axis severity $w^\top\zeta_j$. Here,
$w=(1,0.8,0.7,0.6,0.4,0.3,0.3)$ is fixed. This code-free interface corresponds to the
\armTC{} arm. The learned world-model/value/LCB stack is left deferred.

\paragraph{Self-repair pilot.}
The self-repair pilot is defined as an instrument with no learned component. This instrument
does not inject a new signal. The shared CONFIRM/antidote arms, namely \armB{}, \armCF{},
\armGR{}, and \armTC{}, are reapplied in a live-confirmed search-bound sub-stratum. A unit is
kept only if, when the cached incumbent is rerun, the public-pass fraction still remains
inside the band:

\begin{equation}\label{eq:sr}
\text{keep } u \iff m\in M\ \wedge\ \tfrac{1}{3}\le \hat q_0(u)<1,
\qquad \hat q_0(u)=\frac{n_{\mathrm{passed}}}{n_{\mathrm{total}}} .
\end{equation}

For each arm, true unlock $Y_a(u)$ and pooled count $N_a=\sum_u Y_a(u)$ are computed.
Comparisons are made using the one-sided exact McNemar test in Eq.~\eqref{eq:mcnemar}.
In addition, a $50{,}000$-replicate bootstrap risk-difference confidence interval clustered
by $(\textit{benchmark},\textit{task})$ is used.

\paragraph{EOT-JEPA-RL.}
EOT-JEPA-RL (Execution-Outcome Trace JEPA-RL) is used as a deployed fixed-rule outcome
kernel. For each public test, the candidate's observed value $o_j$ is compared with the
expected value $y_j$, and a mechanical outcome trace is constructed:

\begin{equation}\label{eq:eot-delta}
\delta_j=d(o_j,y_j)=
\begin{cases}
o_j-y_j,&\text{float (}|o_j-y_j|\le \mathrm{atol}_j\Rightarrow \text{pass)}\\[2pt]
(\textit{missing},\textit{extra},\textit{first\_diff},\Delta_{\mathrm{len}}),&\text{container}\\[2pt]
0,&o_j=y_j\\[2pt]
\mathrm{reprdiff}(o_j,y_j),&\text{otherwise.}
\end{cases}
\end{equation}

In the container case, the tuple returns the
\textit{missing}/\textit{extra}/\textit{first\_diff} elements and the length difference
$\Delta_{\mathrm{len}}=\mathrm{len}(o_j)-\mathrm{len}(y_j)$. $\mathrm{atol}_j$ denotes the
per-test absolute tolerance, and $\mathrm{reprdiff}$ denotes the difference at the repr-string
level. Each failing outcome is scored with fixed weights:

\begin{equation}\label{eq:eot-sev}
\begin{aligned}
\mathrm{sev}_j={}&w_{\mathrm{crash}}\mathbf{1}[\text{exception}\lor\text{timeout}]
+w_{\mathrm{type}}\mathbf{1}[\mathrm{type}(o_j)\!\ne\!\mathrm{type}(y_j)]\\
&+w_{\mathrm{small}}\mathbf{1}[\text{small literal}]
+w_{\mathrm{anchor}}\mathbf{1}[\text{anchor}] .
\end{aligned}
\end{equation}

Here, $w=(4,2,1,0.5)$ is used. $\mathrm{sev}_j$ is an EOT-local score independent of the
$\sigma$ primitive in Eq.~\eqref{eq:sev}. The type term is evaluated only when an observed
value is available. The most severe $k_F{=}3$ failures ($F_t$) and $k_A{=}2$ anchors
($A_t$) form a kernel. The deployed NOLEARN rule is

\begin{equation}\label{eq:eot-rule}
a_t=
\begin{cases}
\textsf{EXC},& |F_t|>0\ \wedge\ n_{\mathrm{exc}}/|F_t|\ge \tfrac{1}{2}\\[2pt]
\textsf{OBS\_ANCHOR},& A_t\ne\varnothing\\[2pt]
\textsf{OBS},&\text{otherwise,}
\end{cases}
\qquad a_t=\armB\ \ \text{if } \mathrm{fire}(K_t)=0 .
\end{equation}

The rationale for EOT is defined as a \emph{placebo-distinctness} hypothesis. According to
this hypothesis, the observed content $O$ carries repair-class information beyond the shape
summary $S(O)$, whereas the content-free placebo does not:

\[
I(C_u;O\mid x,S(O))>0
\qquad\text{and}\qquad
I(C_u;O^{\mathrm{shape}}\mid x,S(O))=0.
\]

This statement is used not as proof, but as a falsifiable hypothesis. The learned LCB/JEPA
policy is left deferred.

Each of these instruments is paired with its own content-free shape placebo:
\mbox{FJR-SHAPE,} RIFT-SHAPE, ECK-SHAPE, \armGR{} and EOT-SHAPE. For EOT, an additional
EOT-DECOY control that corrupts the observed value is used. The corresponding verdicts are
reported in Table~\ref{tab:controllers}.

\section{Results}
\label{sec:results}

The experiments were conducted on HumanEval+ and MBPP+ dead task--cell units using three
frozen open-weight code models served through Ollama
(\texttt{qwen2.5-coder:1.5b}, \texttt{qwen2.5-coder:0.5b-instruct}, and
\texttt{deepseek-coder:1.3b}). Each unit was evaluated under five regeneration arms with the
same output-generation budget. Four rounds were run for each arm, and one candidate was
generated in each round. Candidates were first scored by execution against the public test
tier and, if this tier was passed, against the prompt-hidden EvalPlus tier. True unlock was
defined as at least one candidate generated under the same arm passing both tiers together
(Definition~\ref{def:unlock}). Primary analyses were performed on the fixed sample of
$n=290$. Sensitivity analyses were conducted on 60 disjoint units that did not overlap with
the primary sample.

The main five-arm run was conducted on 350 units, formed by the union of the 290 primary units
and 60 disjoint units. In total, 350 units $\times$ 5 arms $\times$ 4 rounds produced
7{,}000 generations, and 1{,}750 transition files were recorded. The primary statistical
family is based on 5{,}800 generations from the fixed sample. The remaining 1{,}200
generations were produced for the disjoint sensitivity sample. An additional 1{,}400 fresh
generations were obtained in the preregistered sixth-arm follow-up. All pre-run manifests were
committed before generation began. The anti-reuse audit episode triggered in the main run is
reported in \S\ref{sec:audit}, and the lookup-scope episode identified during the sixth-arm
follow-up is reported in \S\ref{sec:deconf}. Generation outputs, unit identities, and scores
were not changed during amendments. Original and amended audit verdicts were preserved
separately. Unless the disjoint sample is explicitly specified, the findings below are reported
on the fixed sample of $n=290$.

\subsection{Primary effects and observed blind baseline}
\label{sec:family}

\begin{figure}[t]
\centering
\includegraphics[width=\linewidth]{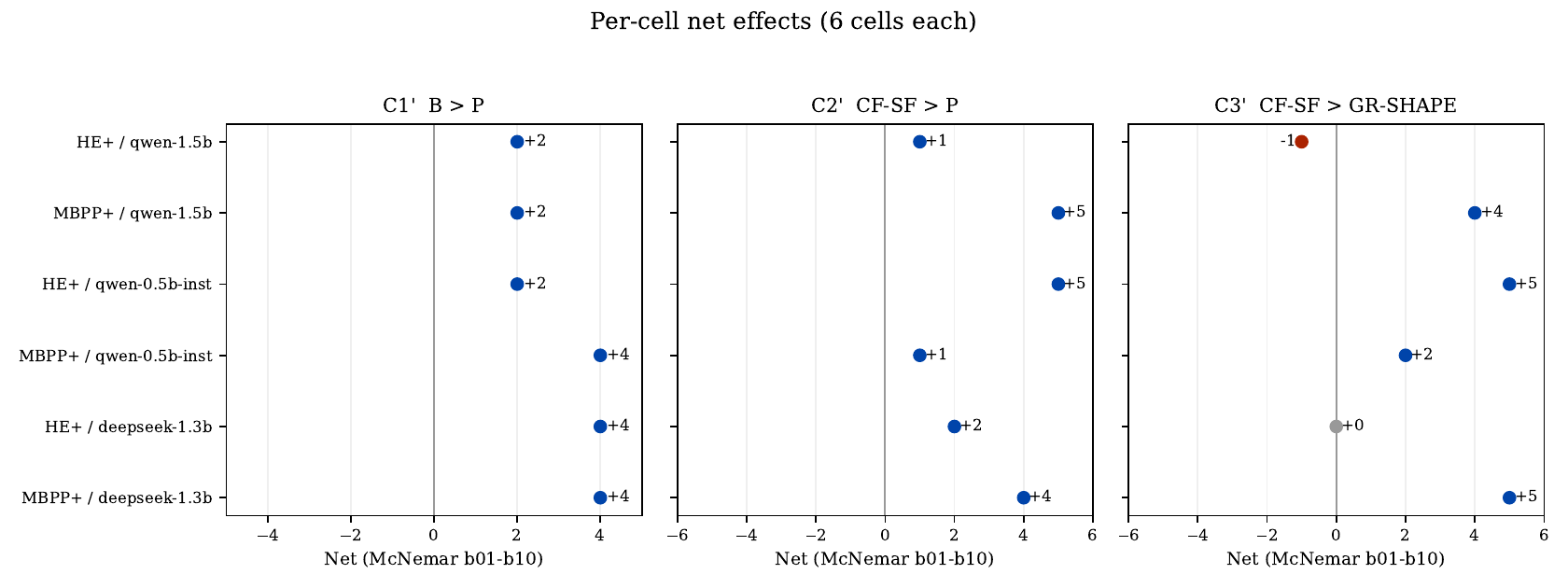}
\caption{Per-cell net discordant effects for the three family claims across the six
benchmark$\times$model cells. The dashed line denotes zero. \Cone{} and \Ctwo{} are in the
positive direction in all six cells. \Cthree{} is positive in four of six cells, with a worst
cell of $-1$.}
\label{fig:percell}
\end{figure}

All three preregistered claims passed the frozen statistical thresholds
(Table~\ref{tab:family} and Figure~\ref{fig:percell}). For the trap hypothesis \Cone{},
blind resampling unlocked 25 units that were not unlocked by the bare-code arm. Seven units
were observed in the reverse direction. Thus, a $+18$ net effect was obtained in favor of
blind resampling (exact one-sided $p=0.00105$, Holm-adjusted $p=0.0021$).

For the antidote hypothesis \Ctwo{}, a $21/3$ discordant split and a $+18$ net effect were
obtained in favor of the code-plus-facts packet
($p=0.000139$, Holm-adjusted $p=0.00042$). Because the two arms followed independent
trajectories, this comparison is reported as an arm-level packet effect rather than as a
same-source-code fact-addition effect.

For the content-attribution claim \Cthree{}, a $22/7$ discordant split and a $+15$ net effect
were observed (Holm-adjusted $p=0.0041$). In this mirror comparison, source code, failure
instruction, template, and bullet count were held fixed, while the fact block was changed
between task-relevant facts and generic bullets. Because exact token length and lexical
distribution were not equalized, this contrast is reported only as content-favoring
component-level evidence under the matched dimensions. Under the frozen statistical decision
rule, the family label was assigned as \textbf{CERTIFIED, content-attributed}. The audit
verdict is reported separately from the statistical label in \S\ref{sec:audit}.

\begin{table}[t]
\caption{Preregistered family on $n=290$ dead units. Pooled exact one-sided McNemar tests were
evaluated within a three-claim Holm family. $\bzo$ denotes units unlocked only by the favored
arm. $\bten$ denotes units unlocked only by the comparison arm. Net effect was computed as
$\bzo-\bten$. The share column reports $\bzo/(\bzo+\bten)$ with the exact Clopper--Pearson
95\% interval. The final column reports the Haldane--Anscombe-smoothed pooled matched-pair
odds ratio.}
\label{tab:family}
\centering\footnotesize\setlength{\tabcolsep}{4pt}
\fitwidth{\input{tables/family}}
\end{table}

The smoothed pooled matched-pair odds ratios computed with Equation~\eqref{eq:orpool} were
$3.4$ for \Cone{}, $6.1$ for \Ctwo{}, and $3.0$ for \Cthree{}. The direction of the
co-primary effects was preserved in all six cells. The preregistered cell guard was defined
for \Cone{} and \Ctwo{} in each benchmark$\times$model cell by the following predicate: a
cell fails the guard if the net discordant difference is $\le -2$ or if the one-sided binomial
$p<0.05$ in the base-only harm direction. No cell triggered this guard. This result indicates
that no conflict was found between the pooled direction and the cell-level signs. It does not
imply that cell-level effect magnitudes were homogeneous or precisely estimated.

\begin{figure}[t]
\centering
\includegraphics[width=\linewidth]{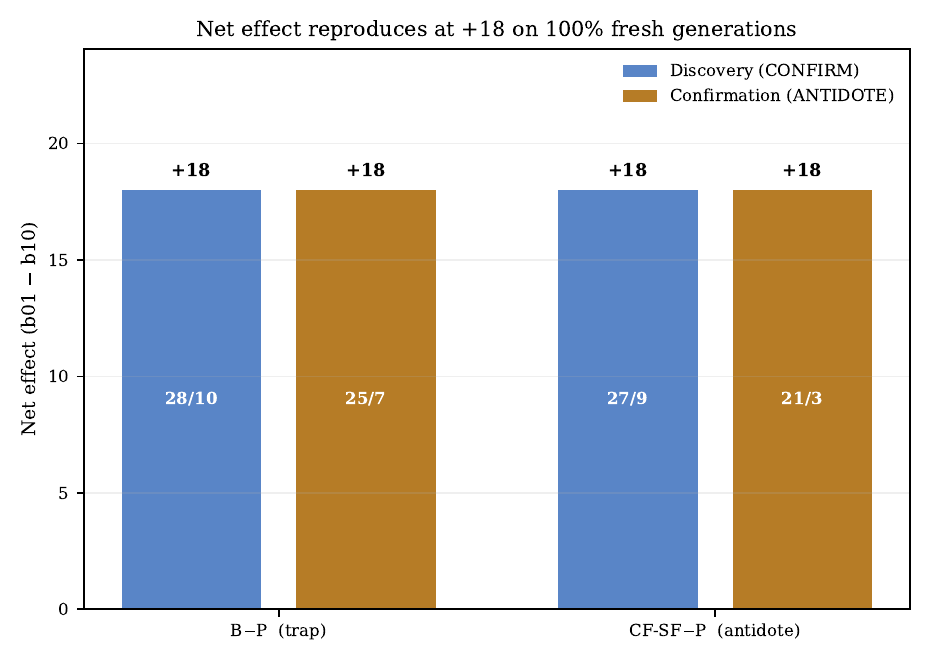}
\caption{Discovery and fresh-generation confirmation. Descriptive contrasts observed with
discovery seeds and preregistered estimates obtained with 100\% fresh confirmation seeds are
shown on the same $n=290$ units. Net effects were reproduced as $+18$ for both contrasts,
although discordant compositions changed.}
\label{fig:reproduction}
\end{figure}

Discovery-stage contrasts were reproduced with the same net magnitude on 100\% fresh seeds
and fresh generations. In the discovery run, $\armB-\armP=+18$ and $\armCF-\armP=+18$ were
observed. In the confirmation run, $+18$ was again obtained for both contrasts
(Figure~\ref{fig:reproduction}). However, discordant compositions changed from $28/10$ to
$25/7$ for the trap contrast and from $27/9$ to $21/3$ for the antidote contrast. Thus,
fresh-generation confirmation shows net-effect agreement on the same fixed sample, not
reproduction of individual task outcomes \citep{gorman2019splits}.

\phantomsection
\label{sec:ladder}

\begin{figure}[t]
\centering
\includegraphics[width=\linewidth]{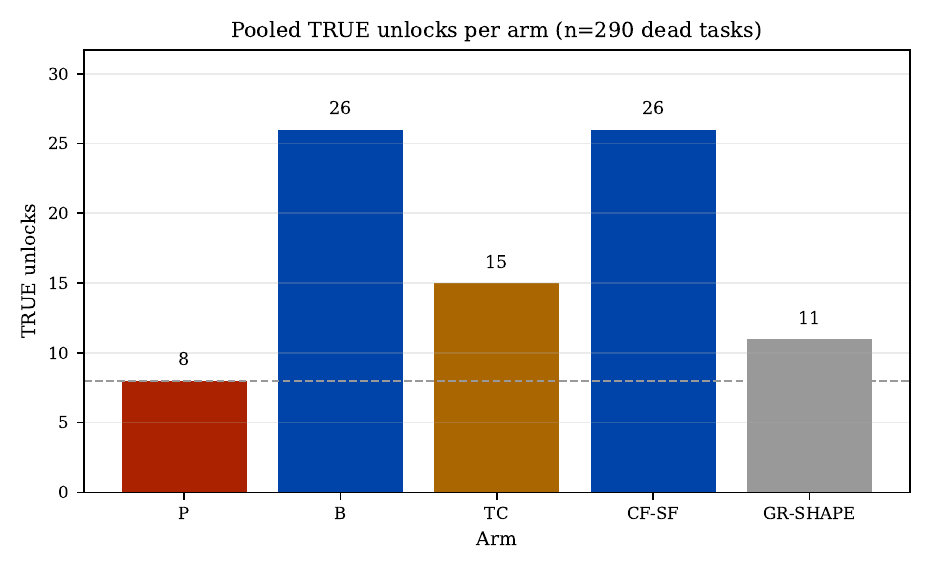}
\caption{Pooled true unlocks across 290 dead units. Blind resampling (\armB) and
code-plus-facts (\armCF) tied with 26 unlocks. Facts only (\armTC) produced 15, placebo
(\armGR) produced 11, and bare code (\armP) produced 8 unlocks.}
\label{fig:unlocks}
\end{figure}

The arm-level performance profile was summarized by pooled true-unlock counts
(Figure~\ref{fig:unlocks} and Table~\ref{tab:percell}). Blind resampling (\armB) and
code-plus-facts (\armCF) produced 26 unlocks each. Facts only (\armTC) produced 15, placebo
(\armGR) produced 11, and bare code (\armP) produced 8 unlocks. The observed ordering was

\[
\armB=26=\armCF>\armTC=15>\armGR=11>\armP=8 .
\]

The corresponding true-unlock rates were $9.0\%$ for \armB{} and \armCF{}, $5.2\%$ for
\armTC{}, $3.8\%$ for \armGR{}, and $2.8\%$ for \armP{}. These marginal counts summarize the
general arm profile and are not used as substitutes for paired discordant contrasts.

The generation-level public-pass profile was broadly consistent with the same ordering
(Table~\ref{tab:armprofile}). Of the 1{,}160 candidates generated per arm, $4.0\%$ passed the
public tier under \armCF{}, $3.9\%$ under \armB{}, $2.3\%$ under \armTC{}, $1.9\%$ under
\armGR{}, and $1.8\%$ under \armP{}.

Public-to-true conversion was $26/42=61.9\%$ for \armB{}, $26/35=74.3\%$ for \armCF{}, and
$8/13=61.5\%$ for \armP{}. Thus, although fewer units passed the public tier under \armCF{}
than under \armB{}, the true-unlock count preserved on the prompt-hidden tier was the same for
the two arms.

First true unlocks were distributed across all four rounds. Round counts were $7/10/4/5$
under \armCF{} and $4/4/7/11$ under \armB{}. Because per-round success rates were low and
counts were limited, these profiles are reported descriptively.

\begin{table}[t]
\caption{Per-arm generation profile on the fixed sample ($n=290$ units and 1{,}160
generations per arm). Prompt length is the cell-size-weighted pooled mean. Public-pass rate is
reported as a generation-level proportion. Public and true unlocks are unit counts. The final
column reports the first true-unlock round.}
\label{tab:armprofile}
\centering\footnotesize\setlength{\tabcolsep}{4pt}
\fitwidth{\input{tables/arm_profile}}
\end{table}

\begin{table}[t]
\caption{Per-cell true unlocks by arm and per-cell net discordants for the three family
claims. The pooled row is exactly equal to the sums of the cell-level columns. The
pooled-equals-cell-sum audit invariant was satisfied with zero mismatch.}
\label{tab:percell}
\centering\footnotesize\setlength{\tabcolsep}{4pt}
\fitwidth{\input{tables/percell}}
\end{table}

\begin{figure}[t]
\centering
\includegraphics[width=\linewidth]{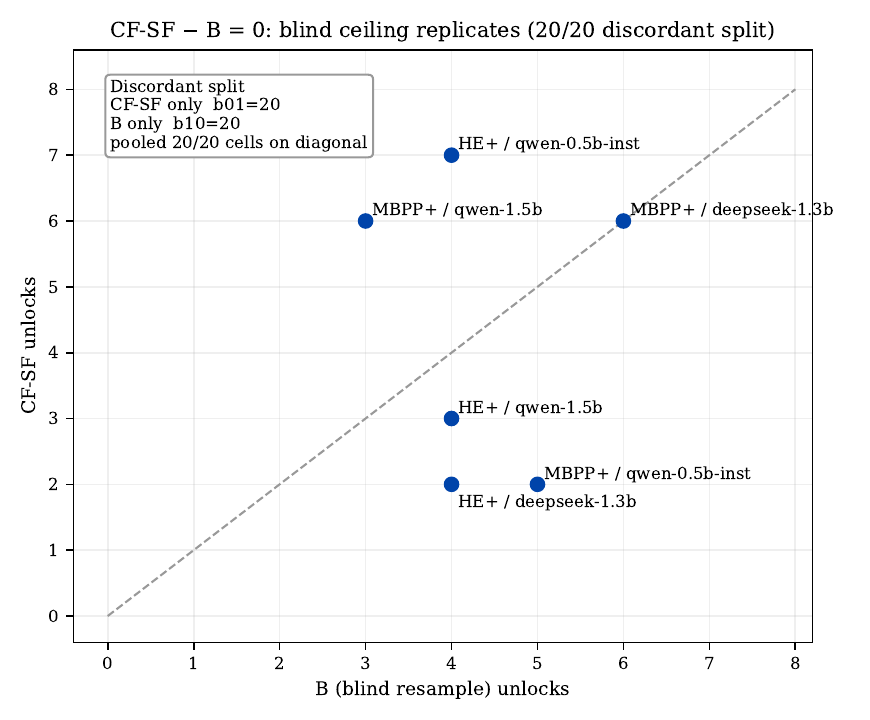}
\caption{Observed comparison with blind resampling. Per-cell true-unlock counts for \armCF{}
and \armB{} are heterogeneous around the identity line. Pooled over all cells, the two arms
tied with 26 unlocks and produced a symmetric $20/20$ discordant split. This result does not
imply equivalence.}
\label{fig:ceiling}
\end{figure}

In the observed comparison with blind resampling, the code-plus-facts arm tied at the pooled
level. Both arms unlocked 26 of the 290 units. However, the unlocked unit sets were not the
same. A symmetric $20/20$ discordant split was observed for the \armCF{}-only and
\armB{}-only directions (Figure~\ref{fig:ceiling}). Thus, the observed pooled difference was
zero, but unit-level agreement was not obtained.

This result does not show equivalence, population-level non-inferiority, or identical
behavior. It only shows that the pooled net difference was zero under the current sample and
output-generation budget.

Per-cell counts were heterogeneous. Blind resampling exceeded code-plus-facts by $1$, $2$,
and $3$ unlocks in three cells. Code-plus-facts exceeded blind resampling by $3$ unlocks in
two cells, and one cell tied. Thus, cell-level differences cancelled within the pooled
average.

The facts-only arm also remained below blind resampling:
$\armTC-\armB=-11$ was obtained ($12/23$, descriptive one-sided $p=0.98$). This descriptive
comparison shows that facts-only conditioning did not produce additional pooled gain over
blind resampling in the tested setting.

\begin{table}[t]
\caption{Standing descriptives and preregistered wording guards ($n=290$ fixed sample).
Descriptive rows are not included in the certified family. The blind-resampling row shows the
observed pooled tie. The material-rescue guard compares placebo rescue with one half of
antidote recovery. The observed-harm gate is a point-estimate wording rule applying the
$\bten/n\le0.02$ threshold and does not provide a safety proof or probabilistic guarantee.}
\label{tab:descriptives}
\centering\small
\fitwidth{\input{tables/descriptives}}
\end{table}

\subsection{Attribution and sensitivity analyses}
\label{sec:deconf}

In the main trap comparison, failing code, failure instruction, and prompt length changed
together. To test the independent contribution of the failure instruction, a sixth arm,
\armBI{} (blind plus instruction), was run under a separate frozen preregistration.
\armBI{} was constructed by adding only the failure instruction used in \armP{} to the
\armB{} scaffold. No code, facts, or bullets were added. Mean prompt length was 406
characters for \armB{}, 478 for \armBI{}, and 1{,}290 for \armP{}.

\armBI{} was run for four rounds on the same 350 units, producing 1{,}400 fresh generations
under a dedicated namespace. Outcomes were paired at the unit level with the \armB{} and
\armP{} outcomes frozen from the main run. Equality audits were applied to model digests,
decoding parameters, and unit lists. The net-count decomposition was

\[
(\armB-\armBI)+(\armBI-\armP)=(\armB-\armP)=+18 .
\]

This equality is reported not as a substantive finding, but as an algebraic bookkeeping
identity showing that the decomposition was correctly constructed on the same units.

No statistically detectable effect was found for the failure instruction. \armB{} exceeded
\armBI{} by $+3$ (D1$^{\prime}$: $18/15$, $p=0.36$). This result does not prove that the
failure instruction has no effect. It only shows that no detectable instruction-only effect
was observed under the current sample and design.

\armBI{} exceeded \armP{} by $+15$ (D2$^{\prime}$: $19/4$, Holm-adjusted $p=0.0026$). The
direction was positive in all six cells, and the smoothed pooled matched-pair odds ratio was
computed as $4.33$. \armBI{} produced 23 total true unlocks and fell between \armB{} and
\armP{} in the pooled ordering.

The preregistered diagnostic follow-up did not detect a statistically distinguishable
instruction-only effect in this sample. The \armBI{}--\armP{} contrast is reported as the
observed diagnostic difference between the code-bearing \armP{} arm and the code-free
instruction arm. Because code content remained confounded with additional prompt length, no
pure code-content causation claim is made (\S\ref{sec:boundaries}). In this follow-up, only
\armBI{} generations were freshly produced, whereas \armB{} and \armP{} outcomes were reused
in frozen form from the main run. Therefore, this analysis is a diagnostic decomposition of
the bare-code contrast, not a fully fresh three-arm confirmatory experiment.

The main and follow-up families were evaluated under separate Holm corrections. When a
conservative Bonferroni correction was applied to all five confirmatory tests across the two
families, \Cone{}, \Ctwo{}, \Cthree{}, and D2$^{\prime}$ remained significant, whereas the
intended-null D1$^{\prime}$ remained non-significant.

\begin{table}[t]
\caption{Sixth-arm family on the same $n=290$ units. Pooled exact one-sided McNemar tests were
evaluated within a two-claim Holm family. \armBI{} generations are fresh, whereas \armB{} and
\armP{} outcomes are frozen from the main run. By construction, the net differences sum to
$+3+15=+18$.}
\label{tab:deconf}
\centering\footnotesize\setlength{\tabcolsep}{4pt}
\fitwidth{\input{tables/deconf_family}}
\end{table}

\phantomsection
\label{sec:robustness}

In the content-attribution and sensitivity analysis, the \armCF{}--\armGR{} mirror contrast
used the same source code, failure instruction, template, and bullet count. Executed facts
were replaced with generic requirement bullets. On the primary sample,
$\armCF-\armGR=+15$ was obtained (Holm-adjusted $p=0.0041$). The descriptive rescue of the
placebo over the bare-code arm was $+3$ ($11/8$, $p=0.32$). This value was below the
preregistered material-rescue threshold of $+9$. This result shows that observed recovery on
the primary sample was in the direction of task-relevant fact content under the matched
dimensions. Because exact token length and lexical distribution were not equalized,
attribution is limited to the matched dimensions defined in the Methods section.

The empty-fact exposure audit showed that, for \armTC{} and \armCF{} in the primary sample,
round-$0$ per-unit fact counts were distributed between $1$ and $6$, and that all $290$ units
received at least one genuine executed fact at round-$0$ entry
(Table~\ref{tab:factexposure}). The rendered violation block is round-specific and is derived
from the incumbent trajectory of each round. Per-round SHA-$256$ equality was preserved across
all four rounds for $250$ of the $290$ units and varied for the remaining $40$ units. Because
no unit received a zero-fact prompt at round-$0$ entry, the post hoc \Cthree{} sensitivity on
units receiving at least one genuine fact is identical to the primary \Cthree{} sample. This
analysis is post hoc and does not generate a new confirmatory claim.

\begin{table}[t]
\centering\footnotesize
\caption{Empty-fact exposure. Round-$0$ rendered violation fact counts for the facts-only
(\armTC) and code-plus-facts (\armCF) arms across the $290$ primary units. The rendered
violation block is round-specific, derived from each round's incumbent trajectory. Per-round
SHA-$256$ equality holds across all four rounds for $250$ of $290$ units and varies for the
remaining $40$ (Round-identical column). No primary unit received zero facts at round-$0$ entry,
although a later round may render zero facts once a candidate stops violating the public tier.}
\label{tab:factexposure}
\fitwidth{\input{tables/fact_exposure}}
\end{table}

The disjoint 60-unit sensitivity sample produced a weaker and mixed pattern. Co-primary
directions were preserved, but net magnitudes fell to $+1$. For \Cone{}, a $2/1$ discordant
split was observed, and for \Ctwo{}, a $3/2$ discordant split was observed. The
content-attribution contrast reversed with $-2$ ($1/3$).

On the same sample, sixth-arm comparisons produced $+4$ ($6/2$) for the code-bearing-context
component and $-3$ ($2/5$) for the instruction comparison. The disjoint sample was small,
drawn by seeded random sampling from unselected eligible units separate from the primary
sample, and underpowered by design. It is therefore not interpreted as independent
confirmation. The direction change in \Cthree{} is nevertheless recorded as a sensitivity
result.

The observed-harm gate was applied separately to the two co-primary comparisons. For \Ctwo{},
the reverse-harm share was

\[
3/290=0.0103\le0.02 .
\]

For \Cone{}, it was

\[
7/290=0.0241>0.02 .
\]

Accordingly, the preregistered wording rule permitted bounded wording for the code-plus-facts
packet relative to the bare-code packet. Wording that universally prefers blind resampling to
bare-code retry was not supported.

This gate is based on a point estimate. It does not provide a confidence bound showing that the
population-level harm rate is below $2\%$, nor does it provide a non-inferiority result or a
deployment safety guarantee. Practical wording is limited to the tested sample, observed
contrasts, and preregistered threshold.

Because the same benchmark task could appear in multiple model cells, a post hoc clustered
sensitivity analysis was conducted on the primary sample. This analysis does not modify the
preregistered exact McNemar family and does not define a new confirmatory family. The cluster
key was taken as benchmark$|$task, and the model was not included in the cluster key. Thus,
observations of the same task under different models were kept within the same cluster, and
the 290 primary units corresponded to 202 task clusters. This valid count of 202 task clusters
is numerically identical to, but conceptually different from, the erroneous 202-row key
collapse reported in the audit trail, where 290 units were merged across cells by bare task
keys.

For each contrast, the paired absolute risk difference was computed in percentage points.
Benchmark composition was preserved by resampling clusters within benchmark, and a
$50{,}000$-replicate cluster bootstrap was used to produce percentile $95\%$ confidence
intervals (CIs), with deterministic seed
$\mathrm{sha256}(\texttt{task-cluster-bootstrap-v1})$. Results are given in
Table~\ref{tab:clustered}. The $95\%$ CIs for the \armB{}$-$\armP{},
\armCF{}$-$\armP{}, and \armCF{}$-$\armGR{} contrasts excluded zero. The interval for the
\armCF{}$-$\armB{} (blind) contrast included zero. Because discordant pairs did not cluster
jointly across model cells, the clustered intervals resembled the unit-level pattern and did
not change the qualitative ordering of the four contrasts. Since this analysis is post hoc, it
does not re-establish the nominal error rate of the preregistered tests.

\begin{table}[t]
\centering\footnotesize
\caption{Post hoc task-cluster sensitivity analysis (does not modify the preregistered
family). Cluster key: benchmark$|$task. $50{,}000$ cluster-bootstrap replicates were resampled
within benchmark. Deterministic seed:
$\mathrm{sha256}(\texttt{task-cluster-bootstrap-v1})$. Risk difference is reported in
percentage points. Values were computed from committed primary result manifests using
\texttt{analysis/task\_cluster\_effects.py} and independently verified as float-identical
using \texttt{analysis/verify\_clustered\_effects.py}.}
\label{tab:clustered}
\fitwidth{\input{tables/clustered_effects}}
\end{table}

In a post hoc $q_0$ stratification, the dead pool was divided into four bins according to
$q_0$, the ratio of the number of public assertions passed by the cached incumbent to the
total number of public assertions: $q_0{=}0$ ($n{=}159$), $0{<}q_0{<}1/3$ ($n{=}22$),
$1/3{\le}q_0{<}2/3$ ($n{=}51$), and $2/3{\le}q_0{<}1$ ($n{=}58$). The total remained
$n=290$. This stratification was not preregistered and does not generate a new confirmatory
claim. In the $q_0{=}0$ stratum, the descriptive difference between \armCF{} and \armB{} was
$-8$. Discordant counts are available in the committed manifests. In the search-bound stratum
($1/3{\le}q_0{<}1$, $n{=}109$), the cached \armCF{}$-$\armB{} difference was $+10$
($p{=}0.0154$). The prospective test of this post hoc discovery slice is reported in
\S\ref{sec:controllers}. These values were derived from committed result manifests, and no
generation output or score was changed.

\subsection{Audit outcomes}
\label{sec:audit}

The sixth-arm evaluator initially produced AUDIT-FAILURE. A total of 120 missing-baseline
errors corresponded to the crossing of 60 disjoint units with two baseline arms. Before any
code change was made, the forensic audit showed that baseline records were present but that
the loader read only the fixed-sample collection. A dated amendment expanded the lookup scope
and preserved the fatal predicate for genuinely absent units. The amendment was gated by two
executable counterexamples requiring that the disjoint baseline be resolved and that an
intentionally deleted baseline produce a fatal failure. Original and amended verdicts were
reported together.

The as-frozen evaluator for the main run initially produced AUDIT-FAILURE. The anti-reuse
invariant defined to detect accidental reuse of discovery outputs was triggered on 28 distinct
(output~SHA-256, prompt~SHA-256) pairs that were also present in discovery records. These
pairs were observed in a total of 34 confirmation records.

Before any code change was made, the forensic audit found no seed-payload collision. Records
with output equality had distinct namespaces and distinct seeds. Prompt equality was confirmed
to be expected by design. Because the same task lists, cached pools, and incumbents were
preserved, some round-0 prompts were byte-identical. Collision incidence was 34 records out of
7{,}000, approximately $0.5\%$.

The arm distribution of colliding records was
$\armP=17$, $\armCF=15$, $\armTC=2$, $\armB=0$, and $\armGR=0$.
This distribution was reported descriptively and was not used for causal or directional
interpretation. Because provenance records were distinct, output-hash equality alone was not
accepted as evidence of cross-run reuse.

With the dated amendment, output-hash equality under intentionally identical prompts was
removed as a fatal predicate and retained as a counted diagnostic. Seed-payload collision,
namespace reuse, and provenance inconsistency were retained as fatal predicates.

The amendment was gated by five executable checks, including a counterexample requiring that a
doctored seed-copy remain a fatal failure and a full real-data re-evaluation. The official
amended verdict was computed only after all checks had been completed. Generation data,
outputs, and scores were not changed. Thus, the assumptions of the evaluation pipeline itself
were also left open to executable criticism, and the original failure was not removed post hoc
from the record.

\begin{table}[t]
\centering
\caption{Verdict status by layer. Statistical verdicts and audit verdicts are reported
separately. Original as-frozen AUDIT-FAILURE statuses are preserved. Dated amendments do not
have the same evidential weight as untouched preregistration.}
\label{tab:verdictstatus}
\fitwidth{\input{tables/verdict_status}}
\end{table}

\section{Discussion}
\label{sec:discussion}

The results obtained in this study answer a narrower and more fundamental question than the
extent to which self-repair feedback is useful for small \emph{frozen code models}: which
kind of information inside the feedback packet carries operational signal, and whether that
signal produces value beyond blind resampling. Three main patterns emerged
jointly. First, the bare-code retry arm remained below blind resampling. Second, the
code-plus-facts packet recovered the bare-code deficit at the pooled level, and the same-code
placebo contrast favored task-relevant fact content on the primary sample. Third, this recovery
did not exceed blind resampling, and code-plus-facts and blind resampling produced zero
observed pooled difference in the tested sample. Therefore, in the tested regime, the observed
function of feedback design was not to produce additional pooled unlocks over blind
resampling, but to close the pooled unlock deficit induced by the code-bearing packet.

This pattern shows that self-repair cannot be interpreted simply as ``the model can correct
itself if it sees its own error.'' Re-showing bare failing code does not provide new external
information about the candidate program. Instead, the previous failed solution is moved back
into the conditioning context. This operation may pull the model's sampling distribution
toward the local trajectory of the failed solution and may narrow the exploration of the
alternative solution space. The preregistered sixth-arm follow-up narrows this interpretation.
Adding the failure instruction without code did not produce a statistically detectable
difference. The remaining difference was associated with code-bearing context. However,
because code content was not separated from intrinsic prompt length, it is not claimed that
failing code alone produced harm. The more limited inference supported by the data is that the
bare code-bearing packet produced lower outcomes than blind resampling and that this difference
was not explained by an instruction-only manipulation.

One plausible explanation for this pattern is anchoring. A concrete but wrong artifact may
pull the generation distribution toward the local neighborhood of the failed solution.
Concrete references have long been known to produce anchoring in human judgment
\citep{tversky1974judgment}. In code models, buggy context has been shown to reduce pass rates
\citep{dinh2023bugs}, and a related-but-wrong function has been shown to anchor generated
outputs in a way that cannot be explained only by verbatim copying
\citep{jones2022cognitivebiases}. Self-preference effects observed in large language model
(LLM) evaluators are also consistent with the possibility that a model may assign
disproportionate weight to its own outputs \citep{panickssery2024selfrecognition}. Nevertheless,
the present experiment does not measure token-level or activation-level mechanisms.
Irrelevant-context distraction, effective-attention reduction, prompt-length burden, and
trajectory persistence could produce the same arm-level pattern. Anchoring is therefore
presented not as a confirmed mechanism, but as a parsimonious explanation consistent with the
measured contrasts.

Executed facts have a different epistemic status. Violation facts are derived not from the
model's verbal self-critique or from its introspective judgment about its own program, but from
the execution of candidate code on public tests. Thus, checker-grounded information, rather
than a model-internal assertion, is added to the feedback packet
\citep{huang2024cannot,olausson2024selfrepair}. It has also been reported that verified
test-case content can be useful for code generation \citep{rizqullah2026tdp} and that
usefulness is associated more with verification status than with mere text presence
\citep{zhang2023algo}. The same-code placebo contrast on the primary sample supports this
distinction as well. Executed facts produce higher outcomes than generic companion text. The
safest interpretation of this result is that the recovery is not explained only by template,
bullet count, or companion-text presence. Rather, under the matched dimensions, recovery is
associated with task-relevant and execution-grounded content. Because exact lexical
distribution and token length are not equalized, all surface-form explanations are not ruled
out. In addition, the direction change of the content-attribution contrast in the disjoint
sensitivity sample limits the external stability of this interpretation.

This distinction carries a direct philosophical consequence for self-repair in LLMs. The
production of a new natural-language critique by an LLM about its own output does not by
itself imply that a strong Popperian refutation has been produced. Such a critique may be
another text artifact generated from the linguistic distribution of the same model. By
contrast, an execution-derived violation is an external comparison that tests the behavioral
claim of the program under an oracle independent of the model. In this study, the generated
program is treated as a tentative \emph{conjecture}, and the executed test violation is
treated as an oracle-relative and executable \emph{counterexample} that refutes a specific
behavioral consequence of that conjecture
\citep{popper1959logic,popper1963conjectures}. The useful signal is obtained not from the
model's use of Popperian vocabulary, not from the model's verbal critique of its own output,
and not from re-exposure to failing code, but from comparing the conjecture with failure facts
revealed by execution. Thus, in the tested regime, falsification functions not as
\emph{exposure}, but as execution-grounded \emph{criticism}.

At the same time, the Popperian framing should not be interpreted here as an instruction that improves model reasoning
through vocabulary alone. The observed effect belongs to executed facts. The Popperian
contribution appears more at the level of measurement architecture. Hypotheses are bounded in
advance, the verdict implied by unfavorable outcomes is specified, discovery outputs are not
reused as confirmation evidence, and the evaluation pipeline itself is kept open to
executable audit invariants. In this sense, the Popperian method functions not as a reasoning
capability possessed by the model, but as a critical-control discipline applied to model
outputs and to the researcher claims produced about those outputs. The methodological
originality lies precisely in this point. Not only the program conjecture, but also the
researcher conjecture that feedback content is useful, is made falsifiable through placebo
controls, a matched-budget baseline, fresh-generation confirmation, and audit
counterexamples.

This two-level structure has direct consequences for LLM evaluation studies. Since LLM
outputs are stochastic, prompt-sensitive, and sensitive to surface form, apparent improvement
may often arise not from content but from prompt length, template, formatting, retry
instruction, or an additional sampling opportunity. Therefore, the apparent success of a
feedback mechanism does not show that it carries epistemically meaningful content.
Falsification is not reduced here to the question of whether the model's answer passed the
test. The questions of which component the improvement can be attributed to, and under which
counterfactual controls the researcher's attribution claim survives, are also made part of
the measurement. For this reason, packet decomposition, placebo mirroring, and matched
output-generation budget are not merely engineering controls. They function as methodological
falsification tools that reduce the distance between claim and evidence in LLM evaluation.

The observed comparison with blind resampling further bounds this interpretation.
Code-plus-facts and blind resampling reached the same pooled unlock total, but they succeeded
on different units. The observed tie showed zero observed pooled difference in the tested
sample. It does not imply equivalence, population-level non-inferiority, or identical
behavior. Nevertheless, none of the tested feedback arms exceeded blind resampling under the
current sample and matched output-generation budget. Code-plus-facts closed the pooled unlock
deficit relative to the bare-code packet, but did not produce additional pooled gain over
blind resampling. This result shows that improvement over the previous completion does not
imply superiority over fresh sampling under the same budget.

In the dead-task regime, the experiment directly measured outcome differences between
conditioning packets. By contrast, the latent generative mechanism that produces these
differences was not identified. In the present design, the tested feedback packets did not
produce observed pooled superiority over blind resampling
\citep{brown2024monkeys,snell2024testtime}. This result makes no claim about the model
distribution or latent support. Whether correct programs exist within the same model
distribution, or which arms would have an advantage under larger sampling budgets, is not the
estimand of this study. What is measured is the paired effect of specific feedback packets on
observed true-unlock outcomes under a finite zero-pass-in-pool regime and a fixed
output-generation budget.

This finding has a broader methodological implication for how inference-time improvement
claims should be evaluated. Methods that exceed a blind baseline may need to modify the search
or generation process more substantively, rather than merely reconditioning on a spent
candidate. Search over diverse plans \citep{wang2025plansearch}, functional decomposition
\citep{chen2024funcoder}, decoding guidance with live execution traces
\citep{lavon2025egcfg}, and inference-aware retraining \citep{chow2025inferenceaware} are
examples in this direction. However, prompt-only variants of this class are prospectively
tested on the search-bound stratum and remain null (Table~\ref{tab:controllers}).
Therefore, under the current evidence, the operational implication is that every richer
feedback claim should be evaluated against blind resampling under an equal output-generation
budget before it is advanced.

When read together with the previous two stages of the falsification-centered measurement
program, a broader signal boundary emerges. In the first stage of the program, Popperian
procedural content was shown not to provide a separable execution-correctness contribution
over a labels-only scaffold, and the measured improvement was associated with scaffold
structure \citep{iscan2026scaffold}. In the same study, a same-model Popperian self-judge did
not exceed random selection. In the second stage of the program, semantic post-hoc
falsification operators, including selection, verification, repair, elimination, and
generation conditioning, were compared against Best-of-$N$ under a deterministic execution
oracle and a leakage-free matched-budget protocol \citep{iscan2026selection}. No semantic
output-space operator exceeded held-out accuracy over Best-of-$N$ in the tested cells.
Measured gains were obtained along axes outside the semantic reasoning space, such as
expression-layer recovery and compute scheduling.

The present study examined the falsification signal within iterative feedback content. This
result does not merely repeat the previous negative findings. Execution-grounded facts
produced measurable recovery relative to a harmful code-bearing packet. However, this recovery
did not produce additional pooled unlocks over blind resampling in the tested sample.
Together, the three stages of the program support the following distinction: Popperian
vocabulary, model-internal self-judgment, and semantic post-hoc reasoning did not carry a
separable accuracy signal in the tested settings. Externally executed counterevidence closed
the pooled unlock gap associated with the code-bearing packet, but did not produce
additional pooled unlocks over blind resampling.

This distinction is central for the epistemology of feedback in LLMs. A critique asserted by
the model and a counterexample produced by an execution oracle are not the same type of
signal. A model-internal critique may be linguistically plausible, but it can remain bound to
the distributional regularities of the same model. An execution-derived counterexample
compares the model's claim with a computational procedure outside the model. Therefore,
epistemic provenance is not merely metadata. The source from which feedback is produced, the
oracle under which it is generated, the auxiliary assumptions on which it depends, and the
placebo controls through which it survives are part of the interpretation of the outcome. In
the tested regime, the useful channel is not model-internal reflection, but externally
grounded error information.

Table~\ref{tab:comparison} positions this contribution against prior work through
experimental controls rather than terminology. The self-repair literature has shown that
execution feedback can be useful on capable or specially trained models
\citep{madaan2023selfrefine,chen2024selfdebug,zhong2024ldb,
zheng2024opencodeinterpreter}. However, because failing code, evidence, and instruction are
usually bundled into a single packet, exposure and evidence cannot be separated. Intrinsic
self-correction critiques have shown that revision can reduce performance when no external
signal is available
\citep{huang2024cannot,valmeekam2023self,stechly2025selfverification}. In these studies,
however, the feedback packet is often removed entirely, so the distinction between the
revision act and prompt content remains unresolved. In the present study, blind
re-generation and bare-code retry are evaluated as separate arms, and lower performance is
shown to be associated more with the code-bearing packet than with the revision instruction.

The matched-budget literature has defined the correct operational baseline and has shown that
repair can be dominated by repeated sampling
\citep{olausson2024selfrepair,brown2024monkeys,snell2024testtime,
zheng2025multiturn}. However, packet ingredients are usually not decomposed with a placebo
control within the same experiment. In this study, same-unit pairing, equal output-sample
counts, prompt-hidden scoring, fresh-generation confirmation, and mirror placebo are combined
within a single instrument. The breadth of the design is deliberately limited: HumanEval+ and
MBPP+, three small models, and a single prompt scaffold are used. In return, packet
components, the operational baseline, and the content placebo are compared under controlled
conditions. This trade-off strengthens internal attribution while limiting external
generalization.

The closest numerical precedents are directionally consistent with the present findings. The
prior result that buggy-context completions fall below no-context completions
\citep{dinh2023bugs} supports the $\armB>\armP$ direction. In the present design, the model's
own failing code was evaluated against a blind operational baseline on the same units. The
result that expert-human feedback improves repair under a matched budget
\citep{olausson2024selfrepair} is consistent with the direction that external evidence can
carry value. Findings that multi-turn feedback can produce small gains or negative outcomes
\citep{zheng2025multiturn} indicate that feedback quantity and useful content must be
separated. The distinct contribution of the present study is the attempt to separate
task-relevant content from matched packet structure through a same-code placebo contrast.

\begin{table}[t]
\caption{Principal experimental controls in the closest prior work and in this study.
``Matched output budget'' means that the same output-generation count was allocated to
feedback and sampling conditions. ``Placebo'' denotes a template-matched packet that contains
no task identifiers or executed facts. ``Fresh generations'' means that newly seeded outputs
were used outside discovery outputs.}
\label{tab:comparison}
\centering\footnotesize\setlength{\tabcolsep}{2pt}
\fitwidth{\begin{tabular}{>{\raggedright\arraybackslash}p{0.20\linewidth} ccccc >{\raggedright\arraybackslash}p{0.24\linewidth}}
\toprule
 & Decomp. & Matched & & Hidden & Fresh & \\
Work & feedback & output budget & Placebo & test & generations & Advantage / limitation \\
\midrule
Self-Refine \citep{madaan2023selfrefine}
& --- & --- & --- & --- & ---
& Broad task transfer, feedback fully bundled \\

Self-Debugging \citep{chen2024selfdebug}
& partial & --- & --- & --- & ---
& Execution traces help, no placebo or budget match \\

LDB \citep{zhong2024ldb}
& --- & --- & --- & --- & ---
& Step-level debugging, capable models and bundled packet \\

OpenCode\-Interpreter \citep{zheng2024opencodeinterpreter}
& --- & --- & --- & --- & ---
& Multi-turn gains, trained model and undecomposed packet \\

CYCLE \citep{ding2024cycle}
& --- & --- & --- & --- & ---
& Improves small models by training, no inference-time isolation \\

DebugBench \citep{tian2024debugbench}
& --- & --- & --- & --- & ---
& Broad debugging benchmark, feedback not decomposed \\

\citet{huang2024cannot}
& --- & \checkmark & --- & --- & ---
& Matched-budget critique, packet removed rather than decomposed \\

\citet{olausson2024selfrepair}
& partial & \checkmark & --- & --- & ---
& Honest budget accounting, bundle compared with baseline \\

\citet{brown2024monkeys}
& --- & \checkmark & --- & \checkmark & ---
& Coverage-scaling baseline, no feedback decomposition \\

\citet{iscan2026scaffold}
& partial & \checkmark & \checkmark & \checkmark & ---
& Labels-only and placebo disambiguation, procedural content does not separate \\

\citet{iscan2026selection}
& --- & \checkmark & partial & \checkmark & ---
& Leakage-free post-hoc evaluation, no semantic operator beats BoN \\

\textbf{This study}
& \checkmark & \checkmark & \checkmark & \checkmark & \checkmark
& Mirror content contrast and fresh-output confirmation, limited scale, language and scaffold \\
\bottomrule
\end{tabular}}
\end{table}

Three bounded practical implications follow from the findings. First, bare failing code should
not be treated as a reliable default feedback packet. In the primary sample, \armP{} produced
the lowest pooled unlock count among the tested arms and was exceeded by blind resampling by
$+18$ net unlocks. However, because the observed-harm gate is not satisfied for the blind
resampling comparison, a universal rule of ``always discard the code'' is not supported. The
supported conclusion is that bare-code retry does not provide reliable superiority over blind
resampling and produces lower pooled outcomes in the current sample.

Second, if failing-code retention is unavoidable, adding executed public-test facts appeared to
be the strongest feedback intervention among the tested options. The code-plus-facts arm
recovered the bare-code deficit at the pooled level and satisfied the observed-harm gate. The
same-code placebo contrast showed that generic diagnostic bullets did not substitute for
executed evidence. This recommendation should be bounded to settings in which source code is
already included in the packet, a public checker is available, and the current model-task
regime applies.

Third, every feedback design should be benchmarked against blind resampling under an equal
output-generation budget. Improvement over the previous completion does not imply superiority
over fresh sampling. In the present experiment, executed facts recovered the bare-code
deficit, but did not exceed the blind baseline. These implications should not be interpreted
as a deployment safety guarantee. The observed-harm gate is based on a point estimate and does
not include a confidence-bound or non-inferiority analysis. In addition, the \armCF{}--\armP{}
comparison is a packet-level contrast involving independent trajectories. Therefore,
practical wording should not state that facts are definitively safe. It should instead state
that, in the tested sample, the code-plus-facts packet produces higher outcomes than the
bare-code packet and ties with blind resampling at the pooled level.

\subsection{Post-primary external-controller follow-ups}
\label{sec:controllers}

After completion of the primary analysis, prospective external-controller experiments were
evaluated on the search-bound stratum to examine whether the primary finding would break under
a more favorable subset and richer feedback channels. These experiments involve prompt-only
controllers carrying joint-embedding predictive architecture (JEPA), reinforcement learning
(RL), and linear upper-confidence-bound (LinUCB) components. The mechanisms of the instruments
are defined in \S\ref{sec:controller-methods}.

\begin{table}[t]
\centering
\caption{Six external-controller experiments conducted on the search-bound stratum
($1/3{\le}q_0{<}1$) after completion of the primary preregistered analysis. All use matched
output budget $R{=}4$ and frozen 0.5--1.5B Ollama models. The endpoint is a fresh candidate
passing both the public and prompt-hidden tiers (true unlock). These experiments were not
preregistered and were not subjected to the audit invariants of the primary frozen evaluation
pipeline. $n$ denotes the number of live-confirmed search-bound units or the specified cell
count. Abbreviations: RD, risk difference; pp, percentage points.}
\label{tab:controllers}
\fitwidth{\input{tables/controller_summary}}
\end{table}

The post hoc $q_0$ stratification (\S\ref{sec:deconf}) showed that the code-bearing packet
produced its highest relative recovery in the search-bound regime. This observation raises the
question of whether external controllers operating on the frozen small model could exceed
blind resampling in that stratum. Six controller experiments were conducted after the primary
study to test this question (Table~\ref{tab:controllers}). None produced a controls-surviving
gain. These experiments are non-preregistered internal follow-ups and are not cited as a
separate reference.

Two experiments failed at the level of realizability before effect measurement. RIFT-JEPA-RL
used a detector operating over ten symbolic inversion axes. Although 19 modules and 136 test
green fixtures were produced, no axis was triggered in any of 24 distinct failing candidates
in the live data (fire-rate $0/24$). Therefore, the controller remained observationally
equivalent to \armB{}. This result shows that green fixtures do not guarantee real live-data
coverage. DCH-JEPA-RL could not be instantiated because the required offline transition corpus
was not available on disk. Thus, it was shown that offline warm-start dependencies must be
verified before the run.

Two experiments produced null or underpowered results. FJR-JEPA-RL (LinUCB + JEPA over prompt
sub-actions) produced FJR$>$\armB{} $=+3$ ($p{=}0.274$) in two cells
($\mathrm{HumanEval}{+}$/qwen2.5-coder-1.5b and 0.5b, $n{=}46$). However, the shape placebo
reproduced the movement with FJR$>$FJR-SHAPE $=-1$ ($p{=}0.773$). This pattern suggests that
the observed movement may be associated more with prompt form than with learned content. The
executable constraint kernel of ECK-JEPA-RL showed interface liveness with a $1.0$ fire-rate.
However, the search-bound cached stratum ($n{=}109$) remained below the $n{\ge}160$ power bar
from the controller's own pre-run power calculation. Since the code-free interface corresponds
to \armTC{}, search-bound \armTC{}$>$\armB{} was limited to $+3$ ($p{=}0.2905$), and the
learned stack was deferred.

The content channels carrying the model's own near-miss code and its own runtime behaviour
were tested directly and prospectively. The self-repair pilot
(\armCF{}: the model's own near-miss code + executable public failing assertions) was run on
69 live-confirmed search-bound units with $1{,}104$ fresh generations. \armCF{}$>$\armB{}
was found to be $+6$ ($p{=}0.073$, cluster-bootstrap risk difference (RD) $+8.7$ percentage
points (pp), 95\% confidence interval (CI) $[-1.4,+18.5]$ pp, including zero). However, the
decisive control \armCF{}$>$\armGR{} was only $+1$ ($p{=}0.50$). The placebo was verified to
be content-free ($0/69$ leakage). The cached post hoc $+10/+11$ discovery was not
prospectively replicated. This result indicates that the discovery may have been a small-$n$
discovery artifact.

EOT-JEPA-RL injected the candidate's actual runtime behaviour as an executable trace. This
channel includes observed-versus-expected values, exceptions, and deltas, and constitutes a
richer content source beyond raw exposure and static tests. On the same 69 units and
$1{,}104$ generations, EOT-NOLEARN$>$\armB{} was $+3$ ($p{=}0.274$), and the decisive
contrast EOT-NOLEARN$>$EOT-SHAPE was $+1$ ($p{=}0.50$). At the full-public-pass level, the
two arms remained equal ($13{=}13$, $n{=}69$). This pattern produced a result almost
identical to the self-repair pilot.

These follow-ups strengthen the philosophical scope of the primary conclusion but do not
change its evidential status. The two strongest content channels in the program, own
near-miss code and own runtime behaviour, prospectively tie with content-free shape placebo
in the most favorable search-bound stratum. Together with the capacity wall in the $q_0{=}0$
stratum, this result points to a practical boundary for gains obtainable through
inference-time content manipulation on a frozen small model. These prospective null results
are consistent with the primary finding: the observed \armCF{}${=}$\armB{} tie and the
\armB{}$>$\armP{} trap in the primary family appear consistent with a form-not-content
ceiling. Exceeding this boundary may require inference-aware retraining or search-space
restructuring rather than frozen-model inference-time manipulation
\citep{chow2025inferenceaware,ding2024cycle}.

\subsection{Validity boundaries and future work}
\label{sec:boundaries}

The validity boundaries of the study should be evaluated across construct, conclusion,
internal, and external validity. Construct validity is strengthened by the execution endpoint
requiring both public and prompt-hidden tests. However, dead-status selection depends only on
the public-tier pass/fail outcomes of the cached pool. The prompt-hidden tier is used only at
the final true-unlock scoring endpoint. Test cases are not shown in any prompt. Nevertheless,
because the prompt-hidden tier defines the final scoring endpoint, it is not called an
untouched external test set. This distinction should be preserved for all claims.

Conclusion validity is supported by exact paired tests, Holm correction, immutable unit keys,
and pre-run power analysis. In contrast, $\widehat{\mathrm{OR}}_{\mathrm{pool}}$ is a pooled
discordant ratio and should not be interpreted as a Cochran--Mantel--Haenszel (CMH) estimator.
No equivalence test is performed for the blind tie. No confidence-bound is computed for the
observed-harm gate. These quantities provide descriptive boundaries.

Internal validity is not equal across contrasts. Since the \armCF{}--\armGR{} contrast holds
source code, failure instruction, template, and bullet count fixed, it is the most tightly
controlled content contrast in the design. However, exact token length and lexical
distribution are not equal. Therefore, the result provides evidence in favor of
task-relevant content, but does not eliminate all surface-form alternatives. In the
\armB{}--\armP{} comparison, failing code, failure instruction, and prompt length change
together. The sixth arm narrows the instruction factor, but does not separate code content
from intrinsic length. A content-free, syntax-matched, and length-matched code-like padding
arm is needed to test the remaining confound.

The \armCF{}--\armP{} comparison is also not a same-code single-ingredient contrast. Whereas
\armCF{} is pinned to the \armTC{} trajectory, \armP{} follows its own trajectory. Therefore,
\Ctwo{} measures the arm-level effect of the code-plus-facts packet relative to the bare-code
packet. It does not measure the isolated effect of adding facts to identical source code.
Same-code content attribution is based primarily on the \armCF{}--\armGR{} contrast. Future
work should include a direct \emph{P+facts} arm that mirrors the \armP{} source code.

The term matched compute is also bounded. Output generation count is equalized. However,
input tokens, wall-clock time, and total floating-point operations (FLOPs) are not equalized.
Longer code-bearing prompts carry higher prefill cost. The design controls the
additional-output advantage, but does not provide strict compute equivalence. Token-matched
and FLOP-matched comparisons would provide a stronger compute claim.

Fresh-generation confirmation is conducted on the same units, cached pools, and incumbents.
This design prevents reuse of discovery outputs and provides seed-level robustness. It does
not provide independent-task replication or benchmark-level generalization. The disjoint
sample preserves the co-primary directions, but the content-attribution contrast reverses.
Therefore, primary \Cthree{} should be considered internally strong but externally unstable.

Preregistration is based on an internal version-control timestamp and is not a third-party
registry. Frozen audit rules produce initial failures during the main and follow-up runs, and
post-run dated amendments are applied. Outputs are not changed, original verdicts are
preserved, and amendments are gated by executable counterexamples. Nevertheless, amended
audit status does not have the same evidential weight as untouched external preregistration.
Independent and externally registered replication would reduce this limitation.

External validity is limited along four main dimensions.

\emph{Model scale.}
Only 0.5B--1.5B frozen models are evaluated. On larger models, the bare-code trap may weaken,
disappear, or produce a different pattern
\citep{olausson2024selfrepair,dinh2023bugs,lam2025codecrash}.

\emph{Task regime.}
The estimand applies to tasks for which no candidate in the finite cached pool passed the
public tier. Effects may shrink or change direction on easier tasks that can be solved by
ordinary sampling.

\emph{Benchmark and language.}
Only Python HumanEval+ and MBPP+ tasks are used. Repository-level, multi-file, interactive,
or different-language coding tasks are not evaluated.

\emph{Prompt scaffold.}
A single prompt template is used for each arm. Robustness over alternative prompt families
and surface forms is not tested.

Future work should directly target the identified boundaries. Priority extensions include a
direct mirror arm that adds executed facts to identical source code from the \armP{}
trajectory, syntax- and length-matched irrelevant-code controls, token- and FLOP-matched
comparisons, externally registered independent-task replication, larger model families,
non-dead tasks, and repository-level benchmarks. Token-level or activation-level analyses are
also needed to distinguish the anchoring account from general distraction, effective-attention
reduction, and trajectory-persistence explanations. This boundary is further supported by the
six prospective external-controller experiments conducted on the most favorable search-bound
stratum (Table~\ref{tab:controllers}). These experiments suggest that, in frozen small
models, inference-time content manipulation may need to be replaced by inference-aware
retraining or search-space restructuring.

Under these boundaries, the central conclusion is bounded as follows. In the primary dead-task
sample, bare code-bearing retry remains below blind resampling. The code-plus-facts packet
recovers this deficit, and the same-code placebo comparison favors executed fact content.
Nevertheless, none of the tested feedback arms exceeds blind resampling, content attribution
is not stable in the disjoint sensitivity sample, and deployment wording is bounded only by
the descriptive observed-harm gate.

The program-level interpretation is broader, but still bounded. In the tested
frozen-small-model settings, the falsification signal does not produce separable accuracy gain
as Popperian vocabulary, model-internal self-judgment, or semantic post-hoc selection.
Execution-grounded counterevidence is able to repair the deficit produced by harmful
conditioning. Therefore, the value carried by falsification here arises not from the model
evaluating its own conjecture more effectively, but from that conjecture being exposed to
externally produced and executable criticism. From the perspective of scientific methodology,
the main contribution is that this critical relation is not left as a philosophical metaphor,
but is embedded into the measurement design. The model conjecture is tested by the execution
oracle, whereas the researcher conjecture is tested through preregistration, placebo controls,
a matched-budget baseline, fresh-generation confirmation, and executable audits. Thus,
falsification is operated in LLM evaluation not merely as a test that filters failed programs,
but as a reflexive methodological discipline that audits the research arrangement producing
evidence, attribution, and claim calibration.

\section{Conclusion}
\label{sec:conclusion}

In this study, the self-repair feedback packet used for small \emph{frozen code models} in
deployment settings where retraining is not feasible was examined at the component level.
Across six benchmark$\times$model cells constructed from HumanEval+ and MBPP+ with three
0.5B--1.5B frozen code models, 290 zero-pass-in-pool task--cell units were evaluated under a
five-arm controlled decomposition. In the main run, $7{,}000$ fresh seeded generations were
produced, and in the preregistered sixth-arm follow-up, $1{,}400$ additional fresh seeded
generations were produced. The output-generation count was equalized across arms, and
correctness was determined by execution on public tests and prompt-hidden EvalPlus tests.
Thus, it was tested directly not only whether self-repair feedback produces more text, a
longer prompt, or an additional retry opportunity, but also which feedback component carries
execution-grounded signal.

The obtained results indicate that bare failing code should not be assumed to be a reliable
feedback object. Blind resampling exceeded the bare-code arm by $+18$ net unlocks
(discordant $25/7$, Holm-adjusted $p=0.0021$). In the preregistered sixth-arm diagnostic
follow-up, no statistically detectable effect was found for the failure instruction, and the
remaining $+15$ difference remained associated with code-bearing context
(Holm-adjusted $p=0.0026$). Nevertheless, because code content was not separated from
intrinsic prompt length, pure code-causation is not claimed. The code-plus-facts packet
produced $+18$ net recovery relative to the bare-code packet (discordant $21/3$,
Holm-adjusted $p=0.00042$). In the same-code placebo contrast, a $+15$ difference was
obtained in favor of executed facts (Holm-adjusted $p=0.0041$). This pattern shows that, on
the primary sample, recovery was associated not merely with generic feedback shape, but with
task-relevant and execution-grounded content under the matched dimensions. However, the
direction change of the content-attribution contrast in the disjoint sensitivity sample forms
a clear boundary for external stability. In addition, because exact token length and lexical
distribution were not equalized, all surface-form explanations were not ruled out.

None of the tested feedback arms exceeded blind resampling under the matched
output-generation budget. Code-plus-facts and blind resampling each produced 26 unlocks, but
did so over different unit sets and showed a symmetric $20/20$ discordant split. This
outcome is not evidence of equivalence, non-inferiority, or identical behavior. It only showed
an observed pooled tie under the current sample and output-generation budget. Therefore, the
main observed function of the code-plus-facts packet was to close the pooled unlock deficit
relative to the bare-code packet, not to produce additional pooled unlocks over blind
resampling. The preregistered observed-harm gate is satisfied for the code-plus-facts versus
bare-code comparison, but is not satisfied for the blind resampling versus bare-code
comparison. This gate should be interpreted only as a point-estimate wording constraint. If
failing code must remain inside the packet, the addition of executed public-test facts is
supported by the current evidence. By contrast, it is not claimed that failing code should
always be discarded, that the code-plus-facts packet is deployment-safe, or that it is superior
to blind resampling.

The philosophical significance of the results is that falsification carries value in LLMs not
as a prompting vocabulary or as a model-internal self-critique capability, but as external and
executable criticism. In this study, the generated program was treated as a tentative
\emph{conjecture}. The violation observed during public execution was evaluated as an
oracle-relative \emph{counterexample} to a specific behavioral consequence of that
conjecture. The findings sharpen when read together with the two preceding
studies, in which Popperian vocabulary, model-internal self-judgment, and semantic post-hoc
falsification did not produce separable accuracy gain in the tested small-model settings.
Externally executed counterevidence was able to repair the loss associated with harmful
conditioning. However, this signal was not observed as a content lever that exceeded blind
resampling. The value carried by falsification here is therefore not the creation of a new
capability in the model, but the opening of the model conjecture to external, auditable, and
execution-grounded criticism.

This boundary is also supported by six prospective external-controller follow-ups conducted
on the search-bound stratum. The two strongest content channels, own near-miss code and
runtime execution trace, tie with content-free shape placebo
(\S\ref{sec:controllers}). These follow-ups should not be treated as preregistered primary
evidence, but as post-primary prospective evidence examining the boundary of the primary
finding in a more favorable stratum. Nevertheless, the observed pattern is consistent with the
interpretation that inference-time content manipulation remains limited in frozen small
models, and that robust gains beyond blind resampling may require inference-aware retraining,
stronger search procedures, or search-space restructuring.

The transferable contribution of the study is not a new repair algorithm. The central
contribution is a reflexive measurement and evaluation methodology that makes claims about
self-repair feedback falsifiable. Packet decomposition, mirror placebo, matched
output-generation budget, same-unit discordant analysis, fresh-generation confirmation,
immutable provenance, and executable audits are combined within a single instrument. Thus,
not only model outputs, but also the evaluation system that produces claims about those
outputs, is left open to critical scrutiny. Preserving original audit failures and dated
amendments in the record prevents negative evidence from being neutralized post hoc and
applies the logic of falsification back onto the research pipeline itself. For this reason,
the final contribution of the study is not that self-repair is improved by a better prompt,
but that LLM feedback claims are transformed into measurable, auditable, and falsifiable
scientific objects under the distinctions of content-vs-form, discovery-vs-confirmation, and
statistical-verdict-vs-audit-verdict.

\section*{Reproducibility and Data Availability}

The research data, result manifests, generated outputs, prompt renders, seed payloads,
audit records, pre-registration and amendment records, generation and evaluation code,
falsification suites, and figure/table generation scripts produced and analyzed in this
study have not been released through a public repository, DOI, or external deposit. These
materials are available from the corresponding author upon reasonable request for academic
verification and replication. Sharing of the requested materials may be limited by file size,
technical transfer constraints, third-party benchmark/model license conditions, and
institutional policy constraints. The open-weight models and EvalPlus benchmarks used in the
study are already available from their respective public sources.

\section*{Ethics and Generative-AI Use Disclosure}

The study evaluates publicly released open-weight code models on public benchmarks. No human
subjects or personal data are involved. Large language models were used in two declared roles:
as the \emph{objects} of the study, namely the three tested frozen models, and as
manuscript-preparation assistants for drafting and editing under the author's direction and
review. All experimental design decisions, pre-registrations, and verdicts belong to the
author. All reported numbers are produced by the executable pipeline described above.

\section*{Funding}

This work was supported by the Scientific and Technological Research Council of Türkiye
(TÜBİTAK) under the 1001 programme, project no. 225M316, “A Tilt-Trirotor Vertical Take-Off
and Landing Controller Enabling Task-Oriented Transfer and Rapid Adaptation:
Hardware-in-the-Loop and Real Validation of a Meta-Learning--Based Reinforcement Learning
Architecture.” The meta-learning code developed in that project was integrated into the test
harness together with the frozen language models studied here and was used during the
agent-design phase to optimize the algorithms designed by the author.

\section*{Conflict of Interest}

The author declares that there is no conflict of interest.

\bibliography{references}

\appendix

\section{Audit and amendment trail}
\label{app:audit}

This appendix records the audit events in the order in which they occurred, together with
their dispositions and evidential status. The purpose is not only to report the final amended
verdict, but also to make the failure modes of the evaluation pipeline itself reproducible and
critically visible. For this reason, the original audit failures were not deleted, dated
amendments were reported separately, and post-amendment verdicts were not substituted for the
original as-frozen verdicts.

\begin{enumerate}[label=(\roman*),leftmargin=2em]

\item \emph{Pre-run audit.}
Before any generation was initiated, the implementation was tested with a falsification suite
containing eight executable counterexamples. The suite covered mirror-equality violation,
seed-namespace violation, foreign-prefix seeds, sample-identity violation, hidden-test
isolation, token-bound violation, resume integrity, and hostile model outputs. The suite
result was recorded as 8/8 passed. Six per-cell pre-run manifests were committed before the
first token was generated. Thus, sample identity, model digest, prompt rendering, and initial
constraints were made version-controlled so that they could not be changed after the run.

\item \emph{Discovery-stage inheritance.}
The pooled-equals-cell-sum invariant and immutable unit keys were inherited from the corrected
evaluator of the discovery stage. In that stage, a pooling-collapse bug was caught by
adversarial review. The bug caused bare task keys to merge units across model cells and
collapsed 290 units into 202 rows. This issue was fixed before the discovery-stage verdict was
issued. In the current study, exact equality between pooled counts and per-cell sums was
preserved as an executable invariant to prevent the same class of error from recurring.

\item \emph{Main post-run anti-reuse audit.}
The as-frozen anti-reuse invariant of the main run returned AUDIT-FAILURE. The failure was
triggered on 28 distinct (output, prompt) hash pairs shared with discovery records. These
pairs appeared in 34 confirmation records in total. Before any code change was made, a
forensic audit was conducted. This audit showed that no seed-payload collision was present,
that the records exhibiting equality had distinct namespaces and distinct seeds, that prompt
equality was expected by design, and that certain round-0 prompts could remain byte-identical
because the same task lists, cached pools, and incumbents were preserved. The record-level
collision incidence was computed as $34/7{,}000 \approx 0.5\%$.

The arm distribution of colliding records was found to be $\armP=17$, $\armCF=15$,
$\armTC=2$, $\armB=0$, and $\armGR=0$. This distribution was recorded descriptively and was
not used for causal or directional interpretation. Because provenance records were distinct,
output-hash equality alone was not accepted as evidence of cross-run reuse.

With the dated amendment, output-pair equality under intentionally identical prompts was
reclassified from a fatal predicate to a counted diagnostic. Seed-payload collision, namespace
reuse, and provenance inconsistency were retained as fatal predicates. The amendment was gated
by a five-counterexample falsification suite, including a counterexample requiring a doctored
seed-copy to remain fatal. The official amended evaluation was recomputed only after the suite
passed 5/5. Both the as-frozen AUDIT-FAILURE verdict and the post-amendment CERTIFIED verdicts
are preserved in the main results documents and in this appendix.

\item \emph{Deconfounding follow-up audit.}
The as-frozen evaluation of the sixth-arm diagnostic follow-up also returned AUDIT-FAILURE.
The failure consisted of exactly 120 missing-baseline errors, corresponding to 60 disjoint
units crossed with two baseline arms. Before any code change was made, the forensic audit
showed that baseline records were present under the disjoint collection of the main manifest,
but that the loader was reading only the fixed-sample collection.

The lookup scope was expanded by a dated amendment. The missing-baseline predicate was
retained as fatal for genuinely absent units. Before the diagnostic follow-up verdict was
computed, the amendment was gated by two added counterexamples requiring the disjoint baseline
to resolve and an intentionally deleted baseline to produce a fatal failure. Original and
amended verdicts are reported separately in \S\ref{sec:deconf}.

\item \emph{Post hoc $q_0$ stratification.}
The primary sample was divided post hoc into four bins according to the cached incumbent's
public-pass fraction $q_0$: $q_0{=}0$ ($n{=}159$), $0{<}q_0{<}1/3$ ($n{=}22$),
$1/3{\le}q_0{<}2/3$ ($n{=}51$), and $2/3{\le}q_0{<}1$ ($n{=}58$). The total was preserved as
$n=290$. This stratification was not preregistered and does not generate a new confirmatory
claim. The calculations were derived from committed result manifests, and no generation output
or score was changed. The corresponding results are reported as post hoc sensitivity evidence
in \S\ref{sec:deconf}.

\item \emph{Prospective controller follow-ups.}
Six external-controller experiments were conducted on the search-bound stratum. Two pilot
runs, namely self-repair and EOT-JEPA-RL, were run on 69 live-confirmed units using separate
namespaces and fresh seeds, with $1{,}104$ generations each. No audit failure was triggered in
either pilot run. These controller experiments were conducted outside the frozen evaluation
pipeline of the primary study. Therefore, they were not subject to the audit invariants of the
primary pipeline. The results and this limitation are reported in \S\ref{sec:controllers}.

\end{enumerate}

This audit trail was kept to show that the statistical verdict and the audit verdict have
different epistemic statuses. Satisfaction of statistical thresholds does not imply that
pipeline-level assumptions are automatically valid. Similarly, the amended verdict obtained
after a dated amendment does not mean that the original as-frozen failure has been removed
from the record. This distinction was preserved so that the evaluation system's own
conjectures would also remain open to executable counterexamples.

\section{Prompt scaffolds}
\label{app:prompts}

The five arms share a single general scaffold family and differ only in the conditioning
block. The templates below are reproduced verbatim from the repository's frozen prompt
builders. \texttt{<TASK PROMPT>} denotes the unmodified benchmark task,
\texttt{<INCUMBENT CODE>} denotes the trajectory's best-so-far failing solution, and
\texttt{<FACT BULLETS>} denotes at most six violated public assertions. Fact bullets were
produced by the frozen no-ellipsis renderer as literal input and expected-output equality
statements and were not elided. The incumbent's observed output was not additionally shown in
the prompt. \texttt{<GENERIC BULLETS>} denotes placebo bullets drawn in order from the frozen
generic list and matched in count to the fact bullets.

\paragraph{\armB{} (blind resample).}
{\small\begin{verbatim}
Write a Python solution for this task.

<TASK PROMPT>

Return your answer in a ```python code block.
\end{verbatim}}

\paragraph{\armBI{} (blind + instruction, the deconfounding sixth arm).}
{\small\begin{verbatim}
Write a Python solution for this task.

<TASK PROMPT>

Your solution failed the tests. Write a corrected, different solution.

Return your answer in a ```python code block.
\end{verbatim}}

\paragraph{\armP{} (bare own code).}
{\small\begin{verbatim}
Write a Python solution for this task.

<TASK PROMPT>

A previous solution is shown below. It did not pass the public tests.

Previous solution:
```python
<INCUMBENT CODE>
```

Your solution failed the tests. Write a corrected, different solution.

Return your answer in a ```python code block.
\end{verbatim}}

\paragraph{\armTC{} (facts only).}
{\small\begin{verbatim}
Write a Python solution for this task.

<TASK PROMPT>

Your implementation must satisfy these input/output requirements:
- <FACT BULLETS>

Return your answer in a ```python code block.
\end{verbatim}}

\paragraph{\armCF{} (own code + the same facts).}
{\small\begin{verbatim}
Write a Python solution for this task.

<TASK PROMPT>

A previous solution is shown below. It did not pass the public tests.

Previous solution:
```python
<INCUMBENT CODE>
```

Your implementation must satisfy these input/output requirements:
- <FACT BULLETS>

Your solution failed the tests. Write a corrected, different solution.

Return your answer in a ```python code block.
\end{verbatim}}

\paragraph{\armGR{} (own code + shape-matched generic bullets).}
The \armGR{} template is identical to the \armCF{} template. Only
\texttt{<FACT BULLETS>} is replaced with \texttt{<GENERIC BULLETS>}. The same source-code
block was preserved and verified against \armCF{} through per-round SHA-256 equality. The
bullet count was also kept identical to \armCF{}. Bullets were drawn in order from the frozen
generic-requirement list:
{\small\begin{verbatim}
- the function returns a value of the documented return type
- the implementation handles the documented input domain
- the result is deterministic for the same inputs
- the function preserves the documented ordering requirements
- edge cases described by the task are handled explicitly
- the implementation follows the task's stated behavior
\end{verbatim}}
An executable forbidden-token audit was applied to every rendered \armGR{} prompt. This audit
verified that no generic bullet contained the task's function name, any public-assert literal
or constant, or any identifier extracted from the task prompt. Within the same (task, round),
the fact bullets shown to the \armTC{} and \armCF{} arms were kept byte-identical. This
equality was enforced by a per-round constraint-block SHA-256 check.

\section{Per-cell unlock rates}
\label{app:rates}

Table~\ref{tab:percellrates} re-expresses the per-cell true-unlock counts in
Table~\ref{tab:percell} as rates for readers who wish to compare cells with unequal sizes.
The pooled row repeats the endpoint percentages reported in \S\ref{sec:ladder}. This table
does not define a new measurement or a new statistical family. Its purpose is to make clearer
how the per-cell counts should be read under different denominators.

\begin{table}[h]
\caption{Per-cell true-unlock rates by arm (percentage of cell units) and the pooled rate over
all 290 units in the final row. Rates are a re-expression of the counts in
Table~\ref{tab:percell}. No new measurement is reported. The pooled ordering is not driven by
a single cell. The per-cell maximum belongs to \armB{} or \armCF{} in five of the six cells.
The only exception is the HE+/deepseek-1.3b cell, where \armTC{} leads with 10.2\%.
\armP{} does not reach the cell maximum in any cell.}
\label{tab:percellrates}
\centering\small
\fitwidth{\input{tables/percell_rates}}
\end{table}

\end{document}

%% file: tables/sample_flow.tex
% Auto-generated by scripts/reconstruct_sample_selection.py (Section F). Do not hand-edit.
\begin{tabular}{llrrrrrr}
\toprule
Benchmark & Model & Initial & Dead & Capped & Excl. & Primary & Disjoint \\
\midrule
HumanEval+ & deepseek-coder:1.3b & 164 & 119 & 60 & 1 & 59 & 15 \\
HumanEval+ & qwen2.5-coder:0.5b & 164 & 40 & 40 & 0 & 40 & 0 \\
HumanEval+ & qwen2.5-coder:1.5b & 164 & 24 & 24 & 0 & 24 & 0 \\
MBPP+ & deepseek-coder:1.3b & 365 & 171 & 60 & 0 & 60 & 30 \\
MBPP+ & qwen2.5-coder:0.5b & 365 & 111 & 60 & 0 & 60 & 15 \\
MBPP+ & qwen2.5-coder:1.5b & 365 & 47 & 47 & 0 & 47 & 0 \\
\midrule
\multicolumn{2}{l}{Total} & 164+365 & 512 & 291 & 1 & 290 & 60 \\
\bottomrule
\end{tabular}

% Initial = effective universe iterated by the deterministic loader. MBPP+ 378 nominal minus 13 loader exclusions (3 order-nondeterministic, 10 unusable I/O) = 365; HumanEval+ 164 unaffected.

%% file: tables/test_tier_provenance.tex
% Auto-generated by scripts/verify_test_tier_provenance.py (Section G).
% Layout hand-tuned for the arXiv single-column manuscript; numbers unchanged.
% Public ($\Ipub$, EvalPlus base input suite) is available to the harness but NOT rendered
% verbatim in any arm; only the <=6 violated public I/O facts appear in TC/CF. Prompt-hidden
% ($\Ihid$, first 300 plus inputs by source order) is shown to no arm. Dead-status is
% public-tier-based; prompt-hidden enters only at final true-unlock scoring. The 300-cap binds
% only on HumanEval+. All tiers use EvalPlus 0.3.1 (pinned in uv.lock).
\small
\setlength{\tabcolsep}{5pt}
\renewcommand{\arraystretch}{1.3}
\begin{tabular}{@{}llcccc@{}}
\toprule
Benchmark & Tier & Source & Prompt exposure & Dead-status & Scoring \\
\midrule
HumanEval+ & public        & EvalPlus $\Ipub$                & $\le 6$ facts (\armTC/\armCF) & \checkmark & \checkmark \\
HumanEval+ & prompt-hidden & EvalPlus $\Ihid$ (first $300$)   & none                          & ---        & \checkmark \\
MBPP+      & public        & EvalPlus $\Ipub$                & $\le 6$ facts (\armTC/\armCF) & \checkmark & \checkmark \\
MBPP+      & prompt-hidden & EvalPlus $\Ihid$ (first $300$)   & none                          & ---        & \checkmark \\
\bottomrule
\end{tabular}

%% file: tables/test_count_summary.tex
% Auto-generated by scripts/verify_test_tier_provenance.py (Section H). Do not hand-edit.
\begin{tabular}{lrrrrl}
\toprule
Benchmark & Tasks & Median plus & Max plus & $>$300 & Subset rule \\
\midrule
HumanEval+ & 164 & 972 & 1000 & 126 & first 300 (source order) \\
MBPP+ & 365 & 105 & 147 & 0 & first 300 (source order) \\
\bottomrule
\end{tabular}

% EvalPlus 0.3.1. The first-300 cap binds only on HumanEval+ (126 tasks truncated, pre-cap max 1000). On MBPP+ the cap never binds (max 147). The cap binds the prompt-hidden tier only. Dead-status selection uses public-tier outcomes only.

%% file: tables/config.tex
% Experiment configuration.
\begin{tabular}{ll}
\toprule
Parameter & Value \\
\midrule
Cells & 6 \\
Models & \texttt{deepseek-coder:1.3b} \\
 & \texttt{qwen2.5-coder:0.5b-instruct-fp16} \\
 & \texttt{qwen2.5-coder:1.5b} \\
Benchmarks & 2 (HumanEval+, MBPP+) \\
Rounds & 4 \\
Arms & 5 (\texttt{P}, \texttt{B}, \texttt{TC}, \texttt{CF-SF}, \texttt{GR-SHAPE}) \\
$n$ (fixed sample) & 290 \\
\texttt{num\_predict} & 1024 (protocol) \\
\texttt{num\_ctx} & 4096 (protocol) \\
Seed namespace & \texttt{antidote\textbar} \\
Total fixed-sample generations & $290 \times 5 \times 4 = 5800$ \\
Disjoint-60 generations & $60 \times 5 \times 4 = 1200$ \\
Main-run total generations & 7000 \\
Sixth-arm follow-up generations & 1{,}400 \\
\bottomrule
\end{tabular}

%% file: tables/family.tex
% Primary pre-registered family (pooled exact one-sided McNemar, Holm over 3, $n=290$).
\begin{tabular}{l r r r r r l r r}
\toprule
Contrast & $b_{01}$ & $b_{10}$ & Net & $p$ (raw) & $p$ (Holm) & Verdict & Share [95\% CI] & $\widehat{\mathrm{OR}}_{\mathrm{pool}}$ \\
\midrule
C1$'$ \quad B $>$ P & 25 & 7 & $+18$ & $0.001051$ & $0.002102$ & \textbf{PASS} & $0.78$ [0.60, 0.91] & $3.40$ \\
C2$'$ \quad CF-SF $>$ P & 21 & 3 & $+18$ & $0.0001386$ & $0.0004157$ & \textbf{PASS} & $0.88$ [0.68, 0.97] & $6.14$ \\
C3$'$ \quad CF-SF $>$ GR-SHAPE & 22 & 7 & $+15$ & $0.004065$ & $0.004065$ & \textbf{PASS} & $0.76$ [0.56, 0.90] & $3.00$ \\
\bottomrule
\end{tabular}

%% file: tables/arm_profile.tex
% Arm profile: weighted means across 6 cells (weighted by cell n_sampled); pass rate is per-generation public pass rate.
\begin{tabular}{l r r r r l}
\toprule
Arm & Mean prompt chars & Pass rate (\%) & Pub unlocks & True unlocks (\% of 290) & Rounds r1/r2/r3/r4 \\
\midrule
P & 1289.9 & 1.8 & 13 & 8 (2.8) & 2/3/2/1 \\
B & 406.4 & 3.9 & 42 & 26 (9.0) & 4/4/7/11 \\
TC & 655.3 & 2.3 & 23 & 15 (5.2) & 2/5/3/5 \\
CF-SF & 1482.9 & 4.0 & 35 & 26 (9.0) & 7/10/4/5 \\
GR-SHAPE & 1475.9 & 1.9 & 15 & 11 (3.8) & 3/4/3/1 \\
\bottomrule
\end{tabular}

%% file: tables/percell.tex
% Per-cell TRUE unlock counts and per-contrast net effects.
\begin{tabular}{lrrrrrrrrr}
\toprule
Cell & $n$ & P & B & TC & CF-SF & GR-SHAPE & C1$'$ & C2$'$ & C3$'$ \\
\midrule
HE+ / qwen-1.5b & 24 & 2 & 4 & 1 & 3 & 4 & $+2$ & $+1$ & $-1$ \\
MBPP+ / qwen-1.5b & 47 & 1 & 3 & 2 & 6 & 2 & $+2$ & $+5$ & $+4$ \\
HE+ / qwen-0.5b-inst & 40 & 2 & 4 & 3 & 7 & 2 & $+2$ & $+5$ & $+5$ \\
MBPP+ / qwen-0.5b-inst & 60 & 1 & 5 & 2 & 2 & 0 & $+4$ & $+1$ & $+2$ \\
HE+ / deepseek-1.3b & 59 & 0 & 4 & 6 & 2 & 2 & $+4$ & $+2$ & $0$ \\
MBPP+ / deepseek-1.3b & 60 & 2 & 6 & 1 & 6 & 1 & $+4$ & $+4$ & $+5$ \\
\bottomrule
\end{tabular}

%% file: tables/descriptives.tex
% Standing descriptives and deployment wording gate.
\begin{tabular}{lrrrrl}
\toprule
Item & $b_{01}$ & $b_{10}$ & Net & $p$ (raw) & Verdict \\
\midrule
CF-SF $-$ B (observed pooled tie) & 20 & 20 & $0$ & $0.5627$ & -- \\
TC $-$ B & 12 & 23 & $-11$ & $0.9795$ & -- \\
GR-SHAPE $-$ P & 11 & 8 & $+3$ & $0.3238$ & -- \\
Rescue guard: GR-SHAPE-P vs (CF-SF-P)/2 & -- & -- & $+3$ vs $9$ & -- & PASS \\
Deployment gate C1' & -- & 7 & 7/290 & 0.0241 & FAIL \\
Deployment gate C2' & -- & 3 & 3/290 & 0.0103 & PASS \\
\bottomrule
\end{tabular}

%% file: tables/deconf_family.tex
% BI-DECONF family (pooled exact one-sided McNemar, Holm over 2, $n=290$; BI generations fresh, B/P baselines frozen from ANTIDOTE).
\begin{tabular}{l r r r r r l r r}
\toprule
Contrast & $b_{01}$ & $b_{10}$ & Net & $p$ (raw) & $p$ (Holm) & Verdict & Share [95\% CI] & $\widehat{\mathrm{OR}}_{\mathrm{pool}}$ \\
\midrule
D1$'$ \quad B $>$ BI & 18 & 15 & $+3$ & $0.3642$ & $0.3642$ & FAIL & $0.55$ [0.36, 0.72] & $1.19$ \\
D2$'$ \quad BI $>$ P & 19 & 4 & $+15$ & $0.0013$ & $0.002599$ & \textbf{PASS} & $0.83$ [0.61, 0.95] & $4.33$ \\
\bottomrule
\end{tabular}

%% file: tables/fact_exposure.tex
% Auto-generated by analysis/fact_trajectory_and_exposure_audit.py (Section 9). Do not hand-edit.
\begin{tabular}{lrrrrrrrrr}
\toprule
Arm & 0 & 1 & 2 & 3 & 4 & 5 & 6 & Zero-fact & Round-identical \\
\midrule
\textsf{TC} & 0 & 43 & 44 & 122 & 23 & 15 & 43 & 0 (0.0\%) & 250/290 \\
\textsf{CF-SF} & 0 & 43 & 44 & 122 & 23 & 15 & 43 & 0 (0.0\%) & 250/290 \\
\bottomrule
\end{tabular}

% Round-identical = units whose rendered fact block has identical SHA-256 across all 4 rounds.

%% file: tables/clustered_effects.tex
% Auto-generated by analysis/verify_clustered_effects.py (Section L). Do not hand-edit.
\begin{tabular}{lrrr}
\toprule
Contrast & Net & RD (pp) & 95\% cluster CI \\
\midrule
$\armB-\armP$ & +18 & +6.21 & $[+2.48, +10.00]$ \\
$\armCF-\armP$ & +18 & +6.21 & $[+3.06, +9.57]$ \\
$\armCF-\armGR$ & +15 & +5.17 & $[+1.68, +8.81]$ \\
$\armCF-\armB$ & +0 & +0.00 & $[-4.40, +4.39]$ \\
\bottomrule
\end{tabular}

% Cluster key (benchmark, task); 202 clusters; 50000 benchmark-stratified replicates; percentile 95\% CI; post hoc.

%% file: tables/verdict_status.tex
% Verdict-status table (Section O). Separates the statistical result from the audit status so
% the two are not conflated. "Post-amendment verdict" is used in place of "official verdict";
% the original as-frozen AUDIT-FAILURE statuses are preserved.
\footnotesize
\setlength{\tabcolsep}{4pt}
\begin{tabular}{p{2.5cm}p{2.0cm}p{3.0cm}p{2.0cm}p{2.0cm}}
\toprule
Layer & As-frozen status & Forensic finding & Dated amendment & Post-amendment status \\
\midrule
Primary statistical family (\Cone, \Ctwo, \Cthree) &
CERTIFIED &
Co-primary direction held in all six cells. No cell guard triggered &
None required &
CERTIFIED (preregistered exact McNemar) \\
\addlinespace
Main-run audit &
AUDIT-FAILURE &
Anti-reuse invariant fired on 34 records / 28 distinct (output, prompt) SHA pairs. Forensic audit found zero seed-payload collision &
Output-pair predicate reclassified as diagnostic. Seed-payload collision kept fatal &
CERTIFIED (suite 5/5) \\
\addlinespace
Diagnostic follow-up (statistical) &
CERTIFIED &
D1$^{\prime}$ instruction-only $+3$, $p=0.36$. D2$^{\prime}$ $+15$, $p=0.0026$ &
None required &
CERTIFIED (two-claim Holm) \\
\addlinespace
Diagnostic follow-up (audit) &
AUDIT-FAILURE &
120 missing-baseline errors (loader read only the fixed-sample collection) &
Lookup widened; missing-baseline kept fatal for genuinely absent units &
CERTIFIED (two added counterexamples) \\
\bottomrule
\end{tabular}

% Evidential limitation: dated amendments do not carry the same evidential weight as the
% untouched preregistration. The statistical verdicts and audit verdicts are reported separately.

%% file: tables/controller_summary.tex
% Six external-controller follow-up experiments over the search-bound stratum.
% Hand-authored (the controller code lives on a separate engineering branch and is
% NOT part of the frozen primary evaluation pipeline). Numbers are post-paper and
% non-preregistered; see Section~\ref{sec:controllers}.
\footnotesize
\setlength{\tabcolsep}{4pt}
\renewcommand{\arraystretch}{1.25}
\begin{tabular}{p{1.6cm} p{2.7cm} c p{2.4cm} p{2.5cm} p{3.0cm}}
\toprule
Controller & Mechanism / injected signal & $n$ & vs \armB{} (net, $p$) & vs placebo (net, $p$) & Verdict / kill-cause \\
\midrule
FJR-JEPA-RL &
LinUCB + JEPA over prompt sub-actions (learned sub-action selection) &
46 &
$+3$ ($p{=}0.274$) &
$-1$ ($p{=}0.773$) &
NULL --- shape placebo reproduced the movement; gain is prompt FORM \\
\addlinespace
RIFT-JEPA-RL &
Symbolic assumption inversion (10 axes) &
--- &
--- &
--- &
NULL --- detector fired on $0/24$ live failing candidates ($\equiv$\armB{}); green fixtures $\neq$ real coverage \\
\addlinespace
DCH-JEPA-RL &
Counter-hypothesis + offline group-disjoint warm-start &
--- &
--- &
--- &
UNREALIZABLE --- required offline transition corpus absent on disk \\
\addlinespace
ECK-JEPA-RL &
Executable constraint kernel (code-free interface $\approx$\armTC{}) &
109 &
\armTC{}$>$\armB{} $+3$ ($p{=}0.2905$) &
--- &
UNDERPOWERED --- $n{<}160$ (own power bar); learned stack deferred \\
\addlinespace
Self-repair pilot &
\armCF{}: own near-miss code + executable public failing assertions &
69 &
$+6$ ($p{=}0.073$); RD $+8.7$ pp [$-1.4,+18.5$] &
vs \armGR{}: $+1$ ($p{=}0.50$); $0/69$ leak &
NULL --- cached $+10/+11$ did not replicate prospectively \\
\addlinespace
EOT-JEPA-RL &
Execution-outcome trace (runtime observed-vs-expected, exceptions, deltas) &
69 &
$+3$ ($p{=}0.274$) &
vs SHAPE: $+1$ ($p{=}0.50$) &
NULL --- near-identical to the self-repair pilot \\
\bottomrule
\end{tabular}

%% file: tables/percell_rates.tex
% Per-cell TRUE unlock rates (\%) per arm, from hidden_true_shares. Pooled row = count/290.
\begin{tabular}{l r r r r r}
\toprule
Cell & P & B & TC & CF-SF & GR-SHAPE \\
\midrule
HE+ / qwen-1.5b & 8.3 & 16.7 & 4.2 & 12.5 & 16.7 \\
MBPP+ / qwen-1.5b & 2.1 & 6.4 & 4.3 & 12.8 & 4.3 \\
HE+ / qwen-0.5b-inst & 5.0 & 10.0 & 7.5 & 17.5 & 5.0 \\
MBPP+ / qwen-0.5b-inst & 1.7 & 8.3 & 3.3 & 3.3 & 0.0 \\
HE+ / deepseek-1.3b & 0.0 & 6.8 & 10.2 & 3.4 & 3.4 \\
MBPP+ / deepseek-1.3b & 3.3 & 10.0 & 1.7 & 10.0 & 1.7 \\
Pooled & 2.8 & 9.0 & 5.2 & 9.0 & 3.8 \\
\bottomrule
\end{tabular}